%% file: main.tex
\documentclass{article}
\usepackage{geometry}
\usepackage{url}
 \usepackage{mathpartir,amssymb,amsthm,mathrsfs,float, mathtools, lscape, stmaryrd}
 \usepackage{authblk}
\usepackage{natbib}
\usepackage{amssymb, stackengine, pifont}
\usepackage{centernot}
\usepackage{pifont}

\newcommand{\xmark}{\text{\ding{55}}}
\stackMath
\usepackage{ifthen}
\usepackage{wrapfig,framed}

\newcounter{sqindex}

\usepackage{enumitem}

\include{macros}

\usepackage{booktabs}   
\usepackage{subcaption} 

\newtheorem{lemma}{Lemma}
\newtheorem{theorem}{Theorem}
\usepackage{xspace}

\begin{document}

\title{Relational Symbolic Execution}         
\author[1]{Gian Pietro Farina}
\author[2]{Stephen Chong}
\author[1]{Marco Gaboardi}
\affil[1]{University at Buffalo, SUNY}
\affil[2]{Harvard University}
\date{}

\maketitle
\begin{abstract} Symbolic execution is a classical program analysis
technique used to show that programs satisfy or violate given
specifications.  In this work we generalize symbolic execution to
support program analysis for relational specifications in the form of
relational properties - these are properties about two runs of two
programs on related inputs, or about two executions of a single
program on related inputs. Relational properties are useful to
formalize notions in security and privacy, and to reason about program
optimizations. We design a relational symbolic execution engine, named
\rse{} which supports interactive refutation, as well as proving of
relational properties for programs written in a language with arrays
and for-like loops.
\end{abstract}

\maketitle

\section{Introduction} Relational properties capture the relations
between the behavior of two programs when run on two inputs, and as a
special case the behavior of one program on two different
inputs. Several safety and security properties can be described as
relational properties: non-interference~\cite{GoguenM82}, ~\cite{GoguenM84},
compiler optimizations~\cite{Benton}, sensitivity and continuity
analysis~\cite{ChaudhuriContinuity,ChaudhuriContinuityRobustness,ReedP10},
and relative cost~\cite{Relcost} are just some examples.

In order to \emph{prove} a relational property, one must ensure that
all the \emph{pairs of related executions} satisfy it, instead of just
single executions.  Similarly, for \emph{finding violations} to
relational properties, we need to find \emph{pairs of related
executions} that violate the property.
A natural way to approach the verification and the testing of
relational properties is through their reduction to standard (unary)
properties through ideas like
self-composition~\cite{Barthe,Terauchi,BartheCrespo} and product
programs~\cite{Barthe-products,Eilers18}.  This approach permits to
use standard program verification and bug-finding
techniques~\cite{Milushev,Hritcu}, and to reduce the problem to
designing convenient and efficient self-compositions and product
programs.

Another way to approach the verification and testing of relational
properties is through relational extensions of standard,
non-relational, techniques for these tasks.  Several works have
explored this approach for techniques such as type
systems~\cite{Barthe2014,Pottier,Nanevski,Barthe2015}, program
logics~\cite{Benton,Barthe2012,Sousa2016}, program
analysis~\cite{KwonHE17}, and abstract
interpretation~\cite{GiacobazziM04,feret:sas01,AssafNSTT17}.  In this
approach, one often aims at giving the user the choice on \emph{how to
explore} the use of the \emph{relational assumptions},
(i.e., relational preconditions, relational intermediate assumptions,
and relational invariants) and a way to relate two programs in
order to prove relational properties. Relational assumptions have a
different flavor than non-relational ones, since they permit to
consider only a subset of the product-relation between inputs, and so
only a subset of the pairs of execution of a program. These are often
the key ingredients for reasoning in a natural way about relational
properties. In this paper, we follow this approach and we propose \emph{relational
symbolic execution} (\rse): a foundational technique combining the
idea of relational analysis of programs and symbolic execution.

\rse{} is a relational symbolic execution
engine for a language with arrays and for-loops.  The
target applications we have in mind are data
analysis and statistics, so we focused on a core calculus which
constitute the basis of languages like R~\cite{R}. In
fact, the design of \rse{} was at an early stage informed by the work
in \cite{Vitek}, on the subset of that language: Core R.
For-loops and arrays provide interesting challenges to
both the design of the operational semantics and to the
representation of the different execution paths in constraints.

\rse{} combines both proving and interactive refutation of relational
properties, with the option of providing loop invariants to
effectively prove or refute properties of programs containing loops.  
\rse is built on a hierarchy of four languages (two relational and two
unary --- two concrete and two symbolic) whose operational semantics are
built on each other in a well-founded manner. In particular, the two
relational languages are based on their unary versions and the
two symbolic languages are, as it usually happens in symbolic execution,
the symbolic versions (i.e., extended with symbolic values) of the concrete ones.
The symbolic operational semantics collect constraints about the
execution of a program, or about pairs of executions of programs, that
can be used to prove or refute relational properties.
This gives the user the
ability to experiment with different ways of proving and interactively
refuting relational properties, e.g., both using a single symbolic
relational execution or using a pair of unary symbolic
executions.

We implemented \rse as a prototype, and we used it for experimenting
with different examples of interactive refutation and verification for
several relational properties coming from different domains. The range
of properties and examples we considered show the flexibility and the
feasibility of our approach.

We also compare \rse with other non-relational methods such as 
self-composition and product programs (which can also be defined using
our tool) in their basic form with no optimization.
We find that our approach, thanks to the use of relational
assumptions, improves in efficiency with respect to
self-composition. Product programs give verification conditions that
are often comparable to the one obtained using relational methods, and
they can use standard symbolic execution tools, but a challenge in
using this technique is the additional cost, in term of design, in
building the product---even if recent developments considerably eased
this task, e.g.,~\cite{Eilers18}. In relational symbolic execution, we do not need
any pre-processing and we can directly analyze a program in a
relational way. This shows a trade-off between the different
techniques which can be exploited accordingly to the concrete target
application.  At the current stage, \rse{} users need to provide
invariants for loops with symbolic guards. We envision for the future
to combine our approach with invariant synthesis techniques,
especially relational ones, e.g., \cite{Qin, Chen, ChenFD17, Sigurbjarnarson18}.

\smallskip

Summarizing, the main contributions of our work are:
\begin{itemize}
\item The design of a relational symbolic execution technique, \rse,
  for a language containing for-loops and arrays. This technique is based on 
  relational and unary symbolic operational semantics that permits to explore the different execution
  paths of programs, maintaining constraints about pairs of executions
  that can be used to prove or refute relational properties.
\item The extension of relational symbolic execution to support
  relational and unary invariants to completely explore a loop with
  symbolic guards.
\item  We have implemented \rse~in a prototype. The implementation
  uses an SMT solver to discharge the generated constraints.
  We show the effectiveness of our approach by analyzing several
  examples for different relational properties. 
\end{itemize}
\noindent \paragraph{Outline} The paper is structured in the following way:
in Section~\ref{sec:informal} we introduce the
different design choices behind \rse~in an informal way. 
Using four running examples Section \ref{sec:examples}
shows at an high level how \rse works and how relational assumptions
help in cutting the search space for proofs and refutation witnesses.
In Section~\ref{sec:conc-languages}, ~\ref{sec:sym-languages}, and  ~\ref{sec:metatheory} we provide the main technical
material describing the four languages behind \rse and the meta
theoretical results that connect them.  Section
\ref{sec:implementation} provides some details about the \rse{}
implementation. In Section \ref{sec:experiments} we provide an experimental
comparison of the relational symbolic approach with other standard
techniques for the verification and bug finding of relational
properties such as self-composition and product programs. Finally, in
Section~\ref{sec:related-works} we discuss related works and in
Section~\ref{sec:conclusion} we conclude by providing a summary of
this work.

\section{Relational symbolic execution: informally}
\label{sec:informal}
In this section, we will give an high-level introduction to the main
characteristics of \rse.
\paragraph{Relational semantics}
\rse is based on a relational operational semantics, which describes
the execution of two, potentially different, programs in two,
potentially different, memories. In this semantics a memory e.g., $\mem$ can
map a variable e.g., $x$, either to a single value, for instance $\mem(x)=4$, or
to a pair of values, for instance $\mem(x)=(3,4)$. In the first case,
we know that in the two executions $x$ will take the same value $4$.
In the second case, $x$ will take two different values in the two  
executions that is 3 and 4. In assertions, when we refer to one of the two executions
of the program we use indexed objects. For instance by writing $x_1$ we
mean the variable $x$ interpreted in the first (left) execution. When we
instead have a precondition that implies that the variable has the same value in both
run we will just avoid indexes and write, for example, just $x$.
The relational character of memories is extended also to the operational semantics of
commands and expressions thanks to a pairing construct
$\pairCmd{\cdot}{\cdot}$.  In the spirit of \cite{Pottier}, with $\pairCmd{c_1}{c_2}$
we denote a pair of commands that might differ in two runs.  These are needed, for
instance, when the guard $e$, of a conditional $\ifte{e}{c_{1}}{c_{2}}$,
evaluates to different values in the two executions, and so the two
executions need to take different branches. For instance, when
evaluating $\ifte{e}{c_{1}}{c_{2}}$, if $e$ evaluates to
$\pairExpr{1}{0}$, the first execution needs to evaluate $c_1$, while
the second one needs to evaluate $c_2$. This situation is resolved  by
using the
command pair $\pairExpr{c_1}{c_2}$. To relationally execute a
paired command $\pairCmd{c_1}{c_2}$ we execute both $c_1$ and
$c_2$ in a unary fashion on two different memories independently and when
they both terminate we merge the two final unary memories in one
final relational memory. 
\paragraph{Symbolic semantics}
To enable symbolic execution, the \rse{} engine also supports symbolic
values $X, Y\dots$ As in standard symbolic execution, a symbolic
value $X$ represents a set of possible concrete values.  However, in
relational symbolic execution, symbolic values can appear also in
pairs $\pairExpr{X}{Y}$.  During the computation, symbolic values are
refined through constraints coming from pre and postconditions,
invariants, and conditionals .  At each step, the constraints describe
all the possible concrete values that symbolic values, and pairs of
symbolic values, can assume.  As a simple example, consider symbolic
execution of the program $\ifte{x=0}{c_{1}}{c_{2}}$ starting with a
memory $\mem$ where $\mem(x)=X$.  Note that the symbolic value $X$
represents an arbitrary concrete value, but the value is the same for
both executions. Symbolic execution of the program would follow both
the first branch (collecting the constraint $X=0$) and the second
branch (collecting the constraint $X\neq 0$).  The two constraints
restrict the set of concrete values that $X$ can represent in the two
branches, respectively.
Consider instead executing the same program but with an initial memory
where $\mem(x)=\pairCmd{X_1}{X_2}$. Here, the two executions map the
variable $x$ to different symbolic values, meaning that the value of
variable $x$ may differ in the two executions. Symbolic execution of
the conditional would generate four possible configurations, based on
all possible combinations of the left and right executions taking the
true and false branches. Using relational assumptions, we can cut the
space of the branches to explore and still get an analysis relational
in nature that allows us to exploit the naturality of this approach
instead of reducing it to a unary approach.
\paragraph{Relational ghost variables} We will make use of
(relational) ghost variables~\cite{MartinPavlola2008} to annotate
programs or to give specifications for them. Ghost variables
are variables that don't correspond to real program entities but appear
only in the specification of a program. For instance when we will
reason about relational cost we will use a relational variable
$\gamma$ which counts the cost of the two runs. Other ghost variables
can be used to reason about other properties for instance covert
channels or trace equivalence. The operational semantics of the languages
does not cover ghost variables by itself, but it can easily be extended
by adding conditions to the rule describing how they evolve during the computation.
For instance when reasoning about cost we can select a (potentially proper)
subset of rules of the semantics which cover the cost model we have in mind, 
and extend them with conditions describing how $\gamma$ evolves.
For simplicity in Section \ref{sec:examples}
we will measure the cost of a program by the number of assignments it performs.
\paragraph{Proving relational specifications}
Throughout the whole paper we will use (relational) Hoare triples to
denote specifications of programs.  That is, we will say that a program
satisfies (or doesn't) the triple $\triple{\Phi}{c}{\Psi}$.  Symbolic execution can be
used to prove valid specifications.  In general, if starting from a
symbolic initial state that satisfies a precondition $\Phi$ we execute
(relationally and) symbolically a program $c$ and we only reach final
states where the path constraints imply the postcondition $\Psi$ we know
that the triple $\triple{\Phi}{c}{\Psi}$ is valid.
\paragraph{Interactive refutation and counterexample generation} The
dual way of reasoning is what symbolic execution is mostly used for.
Symbolic execution searches for final states whose associated path
constraints don't imply the postcondition desired, if they are found
it means that there is at least one state where the desired
postcondition might not hold. Symbolic execution has been proved
useful to generate concrete test cases that demonstrate violation of
specifications. This is usually done by using constraint solvers to
find substitutions for symbolic values that satisfy at the same time
the negation of the postcondition on the final states (the violation
of the specification) and some \emph{path condition} (i.e.,
constraints over symbolic values based on the control flow of the
symbolic execution) guaranteeing the reachability of the
violation. \rse can be used in the same way to find violations of
relational properties.

\paragraph{Loops}
Traditionally, symbolic execution has been used more for bug finding
and testing \cite{King, Khurshid} than for proving.  One of the
reasons for this is that conditionals and loops may create state
explosion, and long (possibly infinite) traces of configurations.  To
improve this situation we extend relational symbolic execution with
loop invariants~\cite{sedbook} so that the symbolic execution of a
loop can be performed by \emph{jumping} over the loop in one step and
by adding an invariant to the path condition.  We design two rules
for unary and relational invariants which allow one to reason in one
step about loops both for proving and for finding counterexamples. 
We will see in Section \ref{sec:examples} that 
using an invariant allows us to reason about arrays with symbolic
length, proving in this way this program satisfies a relational property (Lipschitz continuity)
for arrays of arbitrary length. 
When searching for counterexamples, the situation is a bit more
delicate. Indeed, just providing an inductive invariant may lead to \emph{unrealizable}
counterexamples: satisfiable substitutions that are not produced
by any concrete execution. This can happen when the invariants do not
determine precisely enough the state that can be reached after the
loop.  To avoid this situation in subsection \ref{subsec:strength} we formalize a notion of
\emph{strength} of an invariant. \rse uses this notion to check whether the
invariant provided is strong enough ensuring that if a
counterexample is found, then indeed it corresponds to a concrete
execution (or a pair of concrete executions) violating the (possibly
relational) specification of the program. Using loop with invariants
mitigates in part the state explosion problem but it does not solve it
entirely. A lot of research has focused and still focuses on taming
the state explosion in traditional symbolic execution. These
techniques can also be used for relational symbolic execution in order
to tame this complexity. Since \rse{} is intended as a foundational
work we won't concern ourselves here with integrating the framework
with standard techniques for reducing space explosion, or
loop invariant synthesization, as our goal is to present a different approach
to the verification and interactive refutation of relational properties.

\paragraph{Comparison with self-composition and product programs}
We already discussed how self-composition and product programs are
standard approaches which reduce relational properties to unary
properties, and which allow one to use standard program verification
and bug-finding techniques. At the design level, we do not propose our
approach in contrast with these techniques but as an
alternative. Indeed, one can use \rse also as a standard symbolic
execution engine and use these techniques as a pre-processing phase
transforming the program in its self-composition or product
program. However, we believe that at the technical level, relational
symbolic execution offers, in several situations, some keys advantages
that permit to maximize the relational reasoning.  Indeed in the next
section we will see that we don't need to reason about the functional
correctness of the programs, to prove or disprove (even though to
effectively find counterexamples strong invariants involving a
functional description might be needed) relational properties.  This
property is very useful in relational reasoning since it does allow
one to reduce the complexity of the constraints that one need to
consider. Self-composition cannot directly support this for example
for arrays with symbolic length, while product programs can support it
but it requires more complex invariants than in the case of relational
symbolic execution.  To understand better this kind of trade-offs we
perform an experimental evaluation comparing \rse with
self-composition and product programs in
Section~\ref{sec:experiments}.

\section{Examples}
\label{sec:examples}
In this section we present a few examples for proving and disproving
relational properties of programs. We will hide many details in order
to not distract the reader from the main point of the section which is
to provide a general understanding of the way \rse{} works. For
example, in the following we use assertions and constraints
interchangeably but later on (i.e., Section \ref{sec:conc-languages} and
\ref{sec:sym-languages}) the will be treated differently.
\paragraph{Proving anti-monotonicity of the inverse of cumulative distribution function
  (c.d.f) - concrete bounds}
As a first motivating example we consider the program in Figure
\ref{fig:example-monotone}.  The program takes in input a real number
$q\in[0,1]$ and an array $d$ of size $k\geq 1$ such that $\forall i. 1\leq
i\leq k.  d[i]=P[X\leq i]$, where $X$ is some unspecified random
variable. That is, $d$ represents the c.d.f of a random variable $X$
whose realizations lie in the set $\{1,\dots, k\}$. The program then
proceeds to compute the smallest $x$ such that $P[X\leq x]\geq q$. If
we consider $d$ as its input and $x$ as its output then the program
implements the function $F_{q}^{-1}$, i.e., the inverse c.d.f function.
It is natural to consider the point wise order on c.d.fs
described in Figure \ref{fig:example-monotone}.  The function $F_{q}^{-1}$ then,
obeys the following relational property: $\forall d_1,d_2, q. d_1\preceq_{cdf}
d_2\implies F_{q}^{-1}(d_1)\geq F_{q}^{-1}(d_2)$. This property should hence
be true for the program considered. Let's see how
to see this using \rse{}. \rse{} will start executing the program in
a relational memory with two arrays $d_1$ and $d_2$ with the same length,
say 5 for instance. Every value in the arrays will be
symbolic. These arrays will be related by the following relational
assumption (the precondition) $\Phi\equiv \forall i.1\leq
i \leq 5.\implies d_1[i]\leq d_2[i]$. What we want to show is that in every
final state $x_1\geq x_2$.
At the $i$-th iteration (when $x_h=0$, for $h\in\{1,2\}$) the constraint set
will have the following constraints  $cum_{h}=d_h[1]+\dots+d_h[i-1]$\footnote{Actually it will contain the translation of this assertion
in a constraint, but this is a technical detail.}.  \rse{} has 
now four possible paths to explore given by the outer \textup{{\tt
if-then-else}}, and for three of these there are four others given by the inner one, for a total of
13.  Instead of following a brute force approach and continuing
exploring all the paths we can see that one of the paths is already
unsatisfiable. This because $\Phi$ implies $cum_{2}\geq cum_{1}$ and hence
the path characterized by the constraint $cum_1\geq q \land cum_2< q$
is not satisfiable, and hence not reachable, so it can safely be pruned at
every $i$-th iteration. This pruning was possible thanks to the relational assumption $\Phi$.
Similarly, at every $i$-th iteration, from the symbolic state
characterized by $cum_1\geq q \land cum_2\geq q$ we can
disregard the path with constraints $x_1> 0$ and $x_2\leq 0$. 
Relational reasoning allowed us to reduce the number of paths to
follow at every iteration form 13 to 8. It is easy to see how,
following the remaining paths, \rse{} only reaches final states where
$x_1\geq x_2$ and hence proves the specification.
\begin{figure}
      \[
        \begin{array}{llr}
          1)&\ass{cum}{0};\\
          2)&\ass{x}{0};\\   
          3)&{\tt for}(i\ {\tt in}\ 1:{\tt len}(cdf))\ {\tt do}\\
          4)&\ \ \ \ \ \ \ \ {\tt if} (cum\geq q)\ {\tt then}\\
          5)&\ \ \ \ \ \ \ \ \ \ \ \  {\tt if} (x\leq 0)\ {\tt then}\\
          6)&\ \ \ \ \ \ \ \ \ \ \ \ \ \ \ \ \ \ass{x}{i}\\
          7)&\ \ \ \ \ \ \ \ {\tt else}\\
          8)&\ \ \ \ \ \ \ \ \ \ \ \     \ass{cum}{cum+cdf[i]}\\
    \end{array}
    \]
    \caption{Let $\textbf{CDF}$ the set of c.d.fs.
      The program implements $F_{q}^{-1}:\textbf{CDF}\rightarrow \mathbb{R}$. $F_{q}^{-1}$ is monotonically
      decreasing where the order $\preceq_{cdf}$ on CDF, encoded in finite arrays, is defined as:
    $d_1\preceq_{cdf} d_2 \iff \forall x. d_1[x]\leq d_2[x]$ and we consider the standard order on $\mathbb{R}$.}
  \label{fig:example-monotone}
\end{figure}

\paragraph{Proving k-Lipschitz continuity of sorting - symbolic bounds}
In the second running example - code in Figure \ref{fig:example-lip} -
we will again prove a relational property of a program acting on
arrays. The difference with the previous example is that we will do it
for array of symbolic (arbitrary) size $n$.
To achieve that we will use a very natural relational invariant.
\begin{figure}[h]
      \[
      \begin{array}{llr}
      1)&{\tt for}(i\ {\tt in}\ 1:{\tt len}(a)-1)\ {\tt do}\\
      2)&\ \ \ \ {\tt for}(j\ {\tt in}\ i+1:{\tt len}(a))\ {\tt do}\\   
      3)&\ \ \ \ \ \ \ \ {\tt if} (\arracc{a}{i}>\arracc{a}{j})\ {\tt then}\\
      4)&\ \ \ \ \ \ \ \ \ \ \ \        \ass{z}{\arracc{a}{i}}\\
      5)&\ \ \ \ \ \ \ \ \ \ \ \        \ass{\arracc{a}{i}}{\arracc{a}{j}}\\
      6)&\ \ \ \ \ \ \ \ \ \ \ \        \ass{\arracc{a}{j}}{z}\\
    \end{array}
    \]
  \caption{$k$-Lipschitz continuity of a sorting algorithm.}
  \label{fig:example-lip}
\end{figure}
In general given a sorting algorithm, run on two arrays $a_1, a_2$ of
integers with the same length $n$ and related by the following
relational precondition $\Phi\equiv\forall t.1\leq t\leq n \implies
\lvert a_1[t]-a_2[t]\rvert\leq k$, we expect the sorted arrays
to still satisfy the same condition.
We can see this property as $k$-Lipschitz continuity of a sorting
algorithm with respect to $\ell-$infinity norm in both the input and
output space.  In the program under scrutiny at every iteration of the
inner loop we select the smallest element $a[j]$ in the sub array
$[i+1\dots n]$ and we swap it, if necessary, with $a[i]$. In order to
make sense of this example it's important to understand that the three
lines 4), 5), and 6) which implement the swapping are actually
continuous. Indeed, when \rse{} is executing
the branching instruction, there are four possible ways 
the two executions can proceed, that is: both
take the same branch, or they get different branches.
When the two executions take the same branch then
obviously $\Phi$ still holds.  The following Observation
\ref{lem:lem-abs} guarantees that this is the case also when the two
executions follow different branches.
\begin{observation}
  \label{lem:lem-abs}
  $\forall x, y, z, w, k.\\ |x-y|\leq k, |z-w|\leq k, x>z, y\leq w \implies |z-y|\leq k, |x-w|\leq k$.
\end{observation}
For instance, instantiating $x=\arracc{a_1}{i}, y=\arracc{a_2}{i},
z=\arracc{a_1}{i+1}, w=\arracc{a_2}{i+1}$, ensures that $\Phi$ still
holds when the left execution takes the true branch and the right
execution takes the false branch.
So, omitting synchronization of the loop variables, by using the invariant: $I\equiv \forall t.1\leq
t< i\implies \lvert a_1[t]-a_2[t]\rvert\leq k$ for both the loops we
can \emph{jump} outside of the external loop to a unique state where $I[\len{a}+1/i]$ holds.
This state implies trivially the postcondition.
The important fact to notice here is the very natural invariant that 
relational reasoning allows us to specify. In a unary execution instead
we would have to come up with non trivial invariants allowing us
to prove the functional correctness of the program. We will need
to prove not only that the program produces a sorted sequence but
also that the output is a a 
\paragraph{Refuting cost equivalence - concrete bounds}
In the next example we will use \rse{} to refute a property about a
pair of programs $c_1, c_2$.  Let's consider the programs in Figure
\ref{fig:example-cost}. As we mentioned, \rse{} rules can be extended
to use ghost variables that can be updated at every step of execution
of the abstract machine. We can in this way reason about relational
cost~\cite{Relcost}, by using the relational ghost variable $\gamma$
which gets incremented at every assignment. Let's see this in an
example where the two programs take both in input an array of non
negative symbolic integers of size 5 for instance.  The two programs
would sum in the variable $t$ all their elements up to some value
and save in the variable $o$ the first index in the array that made
$t\geq k$ true.
Obviously the first program has a higher cost in terms of assignments
performed.
We want to refute
that the two programs have the same cost, that is our postcondition to
falsify is $\gamma_1=\gamma_2$, while our precondition would be
$\forall i.1\leq i\leq 5\implies \arracc{a_1}{i}=\arracc{a_2}{i}$.  At
every iteration of the body, for $i$ ranging from 1 to 5, \rse{} would
perform, using a specific rule, one step on the left execution
updating $t$ and no steps in the right execution. So $\gamma_1$ would
be incremented but $\gamma_2$ would not.  Now the two runs are both about to
execute a branching instruction.
If on the left execution the guard is true we perform the
assignment, and the same assignment is performed on the right. Hence
the difference in cost is preserved. If the guard on the left is false we loop,
performing another assignment, while on the second run we don't.
\rse{} would explore these paths finding an initial state, that is
a set of concrete values for the array
for which the execution of the two programs would lead to a final relational state
where $\gamma_1>\gamma_2$. We stress here how \rse{} can, with
specific rules, relationally analyze programs with different
syntactical structures by looking for synchronization points, i.e., branching instructions, to
maximise relational reasoning.
\begin{figure}[h]
\centering
  \begin{minipage}{.25\textwidth}
            \[
        \begin{array}{llr}
          1)&\seq{\ass{t}{0}}{\ass{o}{0}}\\
          2)&{\tt for}(i\ {\tt in}\ 1:{\tt len}(a))\ {\tt do}\\
          3)&\ \ \ \ \ \ \ \ \ass{t}{\arracc{a}{i}+t}\\
          4)&\ \ \ \ \ \ \ \ {\tt if}\ (t\geq k \wedge o\leq 0)\ {\tt then}\\
          5)&\ \ \ \ \ \ \ \ \ \ \ \ \ \ \ \ \ \ass{o}{i}\\
          6)&\\
    \end{array}
    \]
    \caption*{Version 1}
  \end{minipage}%
    \begin{minipage}{0.25\textwidth}
      \[
        \begin{array}{llr}
          &\seq{\ass{t}{0}}{\ass{o}{0}}\\          
          &{\tt for}(i\ {\tt in}\ 1:{\tt len}(a))\ {\tt do}\\
          &\ \ \ \ \ \ \ \ {\tt if}\ (t\geq k)\ {\tt then}\\
          &\ \ \ \ \ \ \ \ \ \ \ \ \ {\tt if}\ (o\leq 0)\ {\tt then}\\
          &\ \ \ \ \ \ \ \ \ \ \ \ \ \ \ \ \ \ass{o}{i}\\
          &\ \ \ \ \ \ \ \ {\tt else}\ \ass{t}{\arracc{a}{i}+t}\\
    \end{array}
    \]
    \caption*{Version 2}
  \end{minipage}
  \caption{The two versions of the program are not cost equivalent.}
  \label{fig:example-cost}
\end{figure}

\paragraph{Refuting non-interference - symbolic bounds with weak invariant}
The next running example involves non-interference\cite{GoguenM82}.
Non-interference was introduced as a strong confidentiality guarantee preventing
information to flow from secret values to public observable values. Non-interference
can be formally stated as a relational property of two executions of a single program
with different inputs: a program $c$ is non-interferent if given two input memories $\mem_1$ and
$\mem_2$ that agree on public data and possibly differ on confidential data, the execution of $c$
on $\mem_1$ and $\mem_2$ results in memories $\mem_1'$ and $\mem_2'$ , respectively, that agree on public data.
That is, secret variables don't interfere with observable public variables.
Let's consider the program $c$ in Figure (\ref{fig:example-ni}).  The
program takes in input a secret vector of integers $s$ and password
vector of integers $p$ of the same length. It then scans the
arrays and checks whether they are point wise equal. If not it saves in
$o$ the index of the first difference.  If we assume
$s$ to be an high level variable and $p, o, t$ low level variables,
this program is obviously interferent. Starting from two memories where
$\alen{s}=\alen{p}\land p_1=p_2$\footnote{Equality on arrays is point wise equality, and
can be easily encoded in a first order logic formula with one
universal quantifier.} we can very well reach a final state where $o_1=o_2 \land t_1 = t_2$ does not hold.
We can check this (i.e., refute non-interference)
for arbitrary length arrays of size $n$.
In particular by using the relational invariant
$I_{w}\equiv (t_1=t_2\land o_1=o_2) \Leftrightarrow s_1=s_2$.
\begin{figure}[h]
      \[
        \begin{array}{llr}
          1)&{\seq{\ass{t}{0}}{\ass{o}{0}}}\\
          2)&{\tt for}(i\ {\tt in}\ 1:{\tt len}(s))\ {\tt do}\\
          3)&\ \ \ \ \ \ \ \ {\tt if} (\arracc{s}{i}\neq \arracc{p}{i} \land o\leq 0)\ {\tt then}\\
          4)&\ \ \ \ \ \ \ \ \ \ \ \        \seq{\ass{o}{1}}{\ass{t}{i}}
    \end{array}
    \]
  \caption{Interferent program}
  \label{fig:example-ni}
\end{figure}
Using \rse{} with that invariant will allow to disprove the
postcondition $\gamma_1=\gamma_2$, but the initial memories that
\rse{} would find might not correspond to real counterexamples this
because the relational invariant was not strong enough.
\paragraph{Counterexample generation for non-interference - symbolic bounds with strong invariant}
In the above program we can get exact counterexamples by
choosing the stronger relational invariant $I_s\equiv
I_w\land t_1=\min_{h}s_1[h]\neq p_1[h] \land t_2=\min_{h}s_2[h]\neq
p_2[h] \land o_1\in\{0,1\} \land o_2\in\{0,1\}$\footnote{Again, this
invariant is expressible in the language, but it can be expressed
easily in the language of our assertions.}. As we can see
we need to specify the functional (unary) behavior of the two programs
in the relational invariant in order to strengthen it. \rse{} would then disprove
the specification by providing a relational initial memory $\mem$ for which the precondition holds
and a final relational memory $\mem'$ related by the operational semantics of the program. For instance:
$\mem(p)=([0],[0]), \mem(s)=([0],[1]), \mem'(o)=(0,1), \mem'(t)=(0,1)$.
\section{Concrete Languages: \lang, \rlang}
\label{sec:conc-languages} 
As already mentioned \rse is composed by four languages.  That is,
we extend the semantics of the simplest language \lang 
in two different directions: relationally (\rlang), and
symbolically (\slang).  And then we extend them both to obtain
\rslang.  In this section we describe the simplest language which is
an imperative language (\lang) that contains for-loops and computes
over integers and arrays of integers, and then extend it to a
relational language (\rlang). We refer to these two languages as
\emph{concrete} to distinguish them from the \emph{symbolic} languages
that we will build on top of them in Section \ref{sec:sym-languages}.

\subsection{\lang}
Programs in \lang have the following grammar, where $v\in\ints$ are values:
\begin{align*}
    e ::=&\ e \oplus e\mid \arracc{a}{e} \mid \len{a} \mid x\mid v\\
    c ::=&\ \skipp\mid \seq{c}{c}\mid \ass{x}{e}\mid \ass{\arracc{a}{e}}{e}\mid \ifte{e}{c}{c} \mid \\
           &\ \cfor{x}{e}{e}{c}
\end{align*}
A variable $x\in\varset$ denotes an integer while an array name
$a\in\vararrset$ denotes a function which maps the set
of natural numbers $\{1, ..., l \}$ to the set $\ints$, with $l$
denoting the length of the array. The set of such functions is denoted by $\carrayset$.
The symbol $\oplus$ denotes an arithmetic operation in $\{+,-,\dots\}$.
Expressions are standardly evaluated using a big step judgment $\econf{\mem}{e} \estep v$ whose
defining rules we omit.  Programs $c$ are evaluated through a, mainly
standard, small step judgment
$\scconf{\mem}{c}\scmstep\scconf{\mem'}{c'}$, where memories
$\mem,\mem' \in \memset$ are partial functions with type
$(\varset\rightarrow \forvalset)\cup(\vararrset \rightarrow
\carrayset)$.  We only show one rule for for-loop construct evaluation
in Figure~\ref{fig:sem-for}.
\begin{figure}
  \begin{mathpar}
    \inferrule[\rulestyle{for-unroll}]
    {
      \econf{\mem}{e_{1}} \estep v_{1}
      \and
      \econf{\mem}{e_{2}} \estep v_{2}
      \and
      v_{1}\leq v_{2}
    }
    {
      \scconf{\mem}{\cfor{x}{e_{1}}{e_{2}}{c}} \scmstep\\\\
\left(\mem, \seq{\ass{x}{v_{1}}}{c}{\tt ;}\,{\tt if\ } v_{2}-v_{1}\begin{array}[t]{@{}l}{\tt
        \ then}\, \cfor{x}{v_{1}+1}{v_2}{c}\cr
       {\tt \ else\ } \skipp\end{array} \right) 
 }
  \end{mathpar}
  \caption{A rule for loops in \lang}
  \label{fig:sem-for}
\end{figure}
Note that for-loops, and thus \lang programs, are always terminating.
\subsection{Assertions, triples, validity}
We state and validate program specifications using Hoare triples 
$\triple{\Phi}{c}{\Psi}$, where $c$ is a command in \lang and $\Phi$ and $\Psi$
(respectively, the pre- and post-condition of the triple) are
assertions. Assertions are first-order logical formulas with primitive predicates that compare
arithmetic expressions $aexp$. The latter are built from expressions in \lang extended
with integer-valued logical variables ($i \in \lvar$) and array
expressions $\alpha$. Array expressions include array names $a$,
and array update expressions $\alpha[aexp_1 \mapsto aexp_2]$, which
denotes the array $\alpha$ with the value at index $aexp_1$ updated to
$aexp_2$. Array expressions allow us to express and reason about
updates on arrays using the the extensional theory of arrays
\cite{mccarthy61}.
The truth of a unary assertion $\Phi$ is evaluated against a memory
$\mem\in\memset$ and a \emph{logical interpretation} $\intlog\in
\intlogs\equiv\ints^\lvar$. We write $\mem\vDash_{\intlog} \Phi$ to
denote that $\Phi$ holds in memory $\mem$ with interpretation
$\intlog$.
The following definition, although standard,
is given because it will later be extended to a relational setting.
\begin{mydef}
  \label{def:u-validity} Let $\Phi$ and $\Psi$ be unary assertions and $c$
be a \lang command.  We say that the triple $\triple{\Phi}{c}{\Psi}$ is
valid, and we write $\vDash\triple{\Phi}{c}{\Psi}$, if and only if 
$\forall \mem_1,\mem_2\in\memset, \intlog\in\intlogs$, if
$\mem_1\vDash_{\intlog} \Phi$ and
$\scconf{\mem_1}{c}\scmsteptr\scconf{\mem_2}{\skipp}$ then
$\mem_2\vDash_{\intlog}\Psi$.
 \end{mydef}
 \subsection{\rlang}
\label{sec:rlang}
 To enable relational reasoning we first build a \emph{relational
  language} \rlang on top of \lang. Intuitively, execution of a
single \rlang program represents the execution of two \lang
programs. Inspired by the approach of \citet{Pottier}, we extend the
grammar of \lang with a pair constructor $\pairCmd{\cdot}{\cdot}$
which can be used at the level of values $\pairCmd{v_1}{v_2}$,
expressions~$\pairCmd{e_1}{e_2}$, or commands $\pairCmd{c_1}{c_2}$.
Notice that $c_i, e_i, v_i$ for $i\in\{1,2\}$ are commands,
expressions, and values in \lang, hence nested pairing is not allowed. This
syntactic invariant is preserved by the rules handling the branching instruction.
Pair constructs are used to
indicate where commands, values, or expressions might be different in
the two unary executions represented by a single \rlang execution.
To define the semantics for \rlang, we first extend memories to allow
program variables to map to pairs of integers, and array variables to
map to pairs of arrays. That is, the type of memories for \rlang
is $(\varset\rightarrow\rforvalset)\cup (\vararrset\rightarrow \rcarrayset)$.
The  semantics of \rlang is defined as a big step judgment
$\reconf{\mem}{e} \restep v$ for expressions and
a small step judgment $\rscconf{\mem}{c} \rscmstep \rscconf{\mem'}{c'}$ for
commands, where $\mem,\mem'$ are relational memories, $c, c'$ are commands in
$\rlang$, $v$ ranges over $\rforvalset$, and $e$ is a relational expression.
Figure ~\ref{fig:sem-rlang} shows a selection of the inference rules for these judgments.
The rules use auxiliary functions $\proj{1}{\cdot}$ and
$\proj{2}{\cdot}$, which project, respectively, the first (left) and
second (right) elements of a pair construct (i.e., $\proj{i}{\pairExpr{c_1}{c_2}}=c_i$,
$\proj{i}{\pairExpr{e_1}{e_2}}=e_i$ with $\proj{i}{v}=v$ when $v\in\ints$),
and are homomorphic for other constructs.
For a relational memory $\mem$, we write $\proj{i}{\mem}$ for the
(unary) memory that projects the co-domain appropriately: $\forall n
\in \dom{\mem}.~\proj{i}{\mem}(n) = \proj{i}{\mem(n)}$. 
Rule \rulestyle{r-lift} is the only evaluation rule for \rlang
expressions. It evaluates the left and right projections of the memory
and expression, and combines the results into either a single value, if
both projections produce the same result, or a pair value otherwise.
Rule $\rulestyle{r-if-false-false}$ shows what happens if the left and
right executions both agree on taking the false branch:
the command $\ifte{e}{c_{\mathit{tt}}}{c_{\mathit{ff}}}$ steps to
command $c_{\mathit{ff}}$. However, if the left and right execution
disagree on which branch to take, we need to introduce a command pair
construct to indicate that the command being executed differs in the
left and right executions. One instance of this is rule $\rulestyle{r-if-false-true}$.
We ensure well-formedness of the paired commands by projecting $c_{\mathit{tt}}$ and
$c_{\mathit{ff}}$ before pairing them up.
Rule $\rulestyle{r-pair-step}$ evaluates a pair command by picking one
projection, non nondeterministically, and evaluating it one step,
using the semantics of \lang. The helper function
$\mergemem{i}{\cdot}{\cdot}$ merges two \lang memories
$\mem_1$ and $\mem_2$ into a \rlang memory, using as few pair values
as possible:
\[
  \mergemem{i}{\mem_{1}}{\mem_{2}}=
  \lambda m.
  \left \{
    \begin{array}{ll}          
      \mem_{1}(m)&\text{if}\ \mem_{1}(m)=\mem_{2}(m)\\
      (\mem_{1}(m),\mem_{2}(m))&\text{otherwise}
    \end{array}
  \right .\    
\]
Another rule, not shown in the figure, reduces
$\scconf{\mem}{\pairCmd{\skipp}{\skipp}}$ to $\scconf{\mem}{\skipp}$.
The rules regarding array assignments now have to take into account
that arrays might differ in the two runs.  In particular, given the
command $\ass{\arracc{a}{e_{l}}}{e_{h}}$ the two expressions $e_l$ and
$e_h$ might evaluate differently in the left and right projections. In
the case where $\mem(a)$ is a unary array but the index expression
evaluates to a pair value then the updated array will be a pair of
arrays, as shown in $\rulestyle{r-arr-ass-split}$.
\begin{figure*}
  \begin{mathpar}
    \inferrule[\rulestyle{r-if-false-false}]
    {\reconf{\mem}{e}\restep \pairExpr{v_{1}}{v_{2}} \\\\ v_{1}\leq 0\and v_{2}\leq 0}
    {\rscconf{\mem}{\ifte{e}{c_{\mathit{tt}}}{c_{\mathit{ff}}}}\rscmstep\rscconf{\mem}{c_{\mathit{ff}}}}

    \inferrule[\rulestyle{r-if-false-true}]
    {
      \reconf{\mem}{e}\restep \pairExpr{v_{1}}{v_{2}}
      \and
      v_{1}\leq 0 \and v_{2}> 0
    }
    { \rscconf{\mem}{\ifte{e}{c_{\mathit{tt}}}{c_{\mathit{ff}}}}\rscmstep
      \rscconf{\mem}{\pairCmd{\proj{1}{c_{\mathit{ff}}}}{\proj{2}{c_{\mathit{tt}}}}}
    }
      
    \inferrule[\rulestyle{r-arr-ass-split}]
    {
      \reconf{\mem}{e_{l}}\restep v_{l}
      \and
      \reconf{\mem}{e_{h}}\restep v_{h}
      \and
      \mem(a)=f
      \\\\
      \proj{1}{v_{l}},\proj{2}{v_{l}}\in\dom{f}
      \and
      \proj{1}{v_{l}} \neq \proj{2}{v_{l}}
      \\\\
      f_{1}=f[ \proj{1}{v_{l}}\mapsto  \proj{1}{v_{h}}]
      \and
      f_{2}=f[ \proj{2}{v_{l}}\mapsto  \proj{2}{v_{h}}]
    }
    {
      \rscconf{\mem}{\ass{\arracc{a}{e_{l}}}{e_{h}}} \rscmstep \rscconf{\mem[a\mapsto \pairExpr{f_{1}}{f_{2}}]}{\skipp}
    }    
    
    \inferrule[\rulestyle{r-pair-step}]
    {
      \{i,j\}=\{1,2\}
      \\\\ 
      \scconf{\proj{i}{\mem}}{c_{i}}\scmstep\scconf{\mem'_{i}}{c'_{i}}
      \\\\
      c'_{j}=c_{j}\\
      \mem'_{j}= \proj{j}{\mem}
      \\\\
      \mem'=\mergemem{i}{\mem'_{1}}{\mem'_2}
    }
    {
      \rscconf{\mem}{\pairCmd{c_{1}}{c_{2}}}\rscmstep\rscconf{\mem'}{\pairCmd{c'_{1}}{c'_{2}}}
    }

    \inferrule[\rulestyle{r-lift}]
    {
      \econf{\proj{1}{\mem}}{\proj{1}{e}}\estep v_{1}
      \\\\
      \econf{\proj{2}{\mem}}{\proj{2}{e}}\estep v_{2}
      \\\\
      v=\left \{
      \begin{array}{rc}
        v_{1} & \text{if}\ v_{1}=v_{2} \cr
        (v_{1},v_{2}) & \text{otherwise}
      \end{array}
    \right .\
  }
  {
    \reconf{\mem}{e}\restep v
  }
  \end{mathpar}
  \caption{Semantics  of  \rlang (selected rules).}
  \label{fig:sem-rlang}
\end{figure*}

\subsection{Relational assertions, relational triples, and relational validity}
\label{sec:relation-assert}
We again use Hoare triples to provide specifications of \rlang
programs.  However, assertions for \rlang must be able to express
properties of both executions of a program, and the relationship
between them. To achieve this, we extend expressions in the language
to include indexed program variables and array variables, that
is we equip an array name $a$ or a program variable $x$
with an index $i\in\{1,2\}$ so that, for
example $a_1$ denotes the array $a$ in the left execution, or $x_2$
denotes the variable $x$ in the right execution.
We refer to the extended language as \emph{relational
assertions}.  We extend
relational operators ($=,\leq,<,\dots$) and binary operators
($+,-,\dots$) to work with two pairs of values in the obvious way, and
adapt the definition of the truth of a relational assertion
($\mem\vDash_\intlog \Phi$) appropriately. Note that logical variables
continue to range only over integers (and not over pairs of
integers). Nonetheless, the logic allows us to express relational and
unary properties easily.  Validity of relational Hoare triples for
\rlang is similar to Definition~\ref{def:u-validity}, except for the
use of relational assertions, relational memories, and the semantic
judgment of \rlang instead of \lang.
In the same spirit of the consistency theorem in \cite{BanerjeeNN16},
the following lemma provides a semantical justification for \rlang{}
with respect to \lang{}.
\begin{lemma}
  Let $\Phi$ and $\Psi$ be relational assertions and
  $c$ be a \lang command. If $~\vDash\triple{\Phi}{c}{\Psi}$ then  for all unary memories $\mem_1,\mem_2,\mem'_1,\mem'_2$
  and for all $\intlog\in\intlogs$
  such that $\mergemem{\cdot}{\mem_1}{\mem_2}\vDash_{\intlog} \Phi$,
  $\scconf{\mem_1}{c}\scmstep^{*}\scconf{\mem'_1}{\skipp}$ and
  $\scconf{\mem_2}{c}\scmstep^{*}\scconf{\mem'_2}{\skipp}$
  then 
  $\mergemem{\cdot}{\mem'_1}{\mem'_2}\vDash_{\intlog} \Psi$.
\end{lemma}
Notice that concrete relational semantics is incomplete with respect
to the unary semantics with respect to traces in the sense that the
iterations of a loop go in lockstep until at least one side terminates
(after which the other side may continue).  In fact, in order to keep
the design of the language simple we only allow pair commands to be
introduced by a branching instruction.  In general this causes
\rlang{} to not be complete with respect to \lang{}.  So it is not
possible to use invariants that hold between different iterations by
using rule such as the \emph{dissonant loop rule} in \cite{Beringer2011}.
Indeed in \rse{} the following two programs cannot not be proved
equivalent, for arbitrary positive $n$:
$\cfor{i}{1}{2*n}{\ass{x}{x+1}}$ and
$\seq{\cfor{i}{1}{n}{\ass{x}{x+1}}}{\cfor{i}{1}{n}{\ass{x}{x+1}}}$.

\section{Symbolic Languages: \slang, \rslang}
\label{sec:sym-languages}

Symbolic execution~\cite{King} extends a language with
\emph{symbolic values} that represent unknown or undetermined concrete
values. Symbolic execution uses symbolic values in logical formulas
that track the conditions under which a particular execution path is
taken. By exploring different execution paths and finding satisfying
assignments to these logical formulas (i.e., finding concrete values
to substitute for symbolic values such that the formulas will be
satisfied), symbolic execution of a program can be used to find
concrete test cases that demonstrate an assertion violation in a
program. Conversely, if all execution paths of a program are explored
and no violation is found, then symbolic execution shows that a
program is guaranteed to meet its specification.  In this section, we
extend the \lang and \rlang languages with symbolic execution, giving
us, respectively, the languages \slang and \rslang. In particular,
\rslang allows us to reason symbolically about two executions of a
\lang program, and thus enables us to look for violations of
relational assertions of \lang programs.  However, we need to define
\slang in order to fully specify the semantics of \rslang, indeed, similarly
to how the semantics of \rlang relies on the semantics of \lang,
the semantics of \rslang relies on the semantics of \slang.

The main insight of symbolic execution is to represent sets of
concrete values (in this case integers) and sets of concrete runs of a
program with symbolic values drawn from a set $\symvalset$.  Symbolic
values can be refined during the computation using constraints
expressed as formulas in some formal theory. For instance, when the
guard $X$ of an {\tt if} construct is symbolic, we might choose to
symbolically execute the true branch and refine the set of possible
concrete values that $X$ denotes by adding the constraint $X>0$ to
the \emph{path condition}.  The connection between symbolic languages
and concrete languages is given by ground substitutions
$\sigma\in\subs\equiv\symvalset\rightarrow\forvalset\cup\carrayset$.
We say that a constraint $\phi$ is satisfiable if there exists a
$\sigma\in\subs$ that makes it true. That is, if substituting all the
symbolic values $X$ appearing in $\phi$ with $\sub{X}{\sigma}$ gives
us a true statement. If that's the case we write $\sigma\models\phi$.
When we are only interested in expressing the satisfiability of
$\phi$ with no interest in specifying the actual substitutions we will
write $\sat{\phi}$.  Given a set of constraints $\cstr$, abusing
notation, we denote by $\cstr$ the constraint
$\displaystyle\bigwedge_{s\in\cstr}s$.
Satisfiable path conditions denote actual concrete executions. That
is, all those concrete executions which assign to the symbolic values
concrete values that make the path condition true. If a path condition
is unsatisfiable then it does not represent any concrete execution.
A set of constraints is \emph{valid} if it is true under
every possible substitution. We denote the validity of a constraint
$\phi$ by $\models \phi$.  Building on the previous section we can now define
the two symbolic languages $\slang$ and $\rslang$.
\subsection{\slang}
\label{sec:sfor}
We extend the syntax of \lang expressions by adding to its values elements
$X\in\symvalset$, denoting symbolic values.  Now memories in \slang
map program variables to either integers or symbolic values. We also
represent symbolic arrays in memory as pairs $(X,v)$, where $v$ is a (concrete
or symbolic) integer value representing the length of the array, and
$X$ is a symbolic value representing the array contents, as in the
standard theory of arrays \cite{mccarthy61}. The content of the
arrays can be refined in a set of constraints described below.
Thus, memories in \slang have the type
$(\varset\rightarrow\sforvalset)\cup(\vararrset\rightarrow\sarrayset)$,
where $\sforvalset\equiv\ints\cup\symvalset$ and
$\sarrayset\equiv\symvalset\times\sforvalset$.
Configurations in \slang are triples $\sscconf{\mem}{c}{\cstr}$ where
$\mem$ is a memory, $c$ is a \slang command, and $\cstr$ is a set of
\emph{constraints}. Constraints are first-order logical formulas with primitive
predicates that compare expressions $(e)$ over concrete ($n\in \ints$),
symbolic values ($X\in\symvalset$) and logical variables ($i\in\lvar$).
Constraint expression $\select{e_1}{e_2}$ represents the
(integer) result of reading the array denoted by $e_1$ at the index
denoted by $e_2$, while $\store{e_1}{e_2}{e_3}$ represents the
(array) result of updating the array denoted by $e_1$ at index $e_2$
with value $e_3$. 
A set of constraints $\cstr$ is used to record restrictions
on symbolic values that must hold in order for program execution to
reach a specific configuration. 

Note that although both assertions and constraints are logical
formulas that include comparisons of expressions, they differ because
assertions may contain program variables and array names but
may not contain symbolic values; constraints on the other hand may
contain symbolic values (including $\select{\cdot}{\cdot}$ and $\store{\cdot}{\cdot}{\cdot}$ expressions) and
may not contain program variables or array variables. Given a memory
$\mem$, we can translate assertions to constraints, using $\mem$ to
replace program variables and array names with the (symbolic or
concrete) values $\mem$ maps them to. We write $\tocstr{\mem}{\cdot}$
for this translation function defined inductively on the shape of the
expression.  Symbolic values can now appear in expressions, so a for
loop executed by unrolling might not terminate.  For this reason
we extend the category of commands to also contain commands
of this form: $\cforinv{x}{e_1}{e_2}{c}{I}$.  Where $I$ is an
assertion intended to be a loop invariant. The two kinds (with and without invariant)
of for-loops are treated as distinct syntactic forms.

The semantics of \slang is defined through a big-step judgment,
$\seconf{\mem}{e}{\cstr} \sestep \send{v}{\cstr'}$, for
expressions, and a small-step judgment
$\sscconf{\mem}{c}{\cstr}\sscmstep \sscconf{\mem'}{c'}{\cstr'}$ for
commands. Figure~\ref{fig:sem-sfor} shows some selected rules defining
the judgments.  Notice that evaluating an expression might generate new symbolic
values, and this is why also $\sestep$ returns an updated set of constraints $\cstr'$.
In rules for conditionals, like the rule \rulestyle{s-if-true}, we
record in the constraint set the information about the control flow path.
Rules handling the conditionals make the small-step operational
semantics non-deterministic, since we have to consider both the
case when the guard reduces to a value greater than 0 and when it
reduces to a value less or equal than 0.
In rules for arrays, we record in the constraint set the description of
arrays. 
For example, the rule \rulestyle{s-arr-read} records the selection in
the constraint using a fresh symbol $Y$ which has never occurred in
the computation before that point. Rule \rulestyle{s-arr-write} evaluates
the index of the array to update and the right hand side of the assignment
after updating the memory it records the array update in the set of constraints.
\begin{figure}
  \begin{mathpar}
    \inferrule[\rulestyle{s-arr-read}]
    {
      \seconf{\mem}{e}{\cstr}\sestep \send{v_{s}}{\cstr'}
      \and
      \mem(a)=(X,v'_{s})
      \and
      \fresh{Y}
    }
    {
      \seconf{\mem}{\arracc{a}{e}}{\cstr}
      \sestep
      \send{Y}{\cstr'\cup\{Y=\select{X}{v_{s}}, v_{s}>0, v_{s}\leq v'_{s}\}}
    }

    \inferrule[\rulestyle{s-arr-write}]
    {
      \seconf{\mem}{e_{1}}{\cstr}\ssebstep\send{v_1}{\cstr'}
      \and
      \seconf{\mem}{e_{2}}{\cstr'}\ssebstep \send{v_2}{\cstr''}
      \\\\
      \mem(a)=(X,v_l)
      \and
      \fresh{Y}
      \and
      \mem'=\mem[a\mapsto(Y, l)]
      \\\\
      \cstr'''\equiv \cstr''\cup\{Y=\store{X}{v_1}{v_2}, v_1>0, v_1\leq l\}
    }
    {
      \sscconf{\mem}{\ass{\arracc{a}{e_{1}}}{e_{2}}}{\cstr}
      \sscmstep
      \sscconf{\mem'}{\skipp}{\cstr'''}
    }
    
    \inferrule[\rulestyle{s-if-true}]
    {
      \seconf{\mem}{e}{\cstr} \sestep \send{v_{s}}{\cstr'}
    }
    {
      \sscconf{\mem}{\ifte{e}{c_{\mathit{tt}}}{c_{\mathit{ff}}}}{\cstr} \sscmstep \send{c_{\mathit{tt}}}{\cstr'\cup\{v_{s}> 0\}}
    }

    \inferrule[\rulestyle{s-for-inv}]
    {
      \seconf{\mem}{e_{1}}{\cstr}\sestep\send{v_{1}}{\cstr'}
      \and
      \seconf{\mem}{e_{2}}{\cstr'}\sestep\send{v_{2}}{\cstr''}\\\\ e_{1},e_{2}\in aexp\and 
      \vDash \triple{I \wedge e_{1}\leq x\wedge x\leq e_{2}}{c_{b}}{I[x+1/x]}
      \\\\
      \vspace{15pt}
      \mem_{f}=\lambda n.\left \{
          \begin{array}{rcl}
          v_{2}, && \text{if}\ n=x \cr
          X, && \text{if}\ n\in\updateC{c_{b}}, n\in\varset,  \fresh{X}\cr
          (X, l), && \text{if}\ n\in\updateC{c_{b}}, \mem(n)=(Z,l), \fresh{X} \cr
          \mem(n) \ && \text{otherwise}          
        \end{array}
      \right .\ 
      \\\\
      \models \cstr''\implies\tocstr{\mem}{I[e_{1}/x]\land e_{1}\leq e_{2}}
      \\\\
      \cstr_{f}=\cstr''\cup\{\tocstr{\mem_{f}}{I[e_{2}+1/x]}\}
    }
    {
      \sscconf{\mem}{\cforinv{x}{e_{1}}{e_{2}}{c_{b}}{I}}{\cstr}\sscmstep\sscconf{\mem_{f}}{\skipp}{\cstr_{f}}
    }
  \end{mathpar}
    \caption{Semantics of \slang (selected rules).}
    \label{fig:sem-sfor}
  \end{figure}
As already mentioned we allow the user to specify invariant for loops
and use the rule \rulestyle{s-for-inv}. This rule allows to skip in
one step the whole unrolling of the for-loop provided that the user
has specified an actual inductive invariant. Specifically, the
semantic judgment $\vDash \triple{I \wedge e_{1}\leq x\wedge x\leq
e_{2}}{c_{b}}{I[x+1/x]}$ imposes that $I$ holds before and after every
iteration of the body of the loop provided that the counter variable
$x$ is between the bounds.  Checking that $e_{1},e_{2}\in aexp$ makes
sure that the premise of the triple is actually an assertion and does
not contain symbolic values, as it could be the case since $e_1, e_2$
are expressions in \slang.  The additional check,
$\models \cstr''\implies\tocstr{\mem}{I[e_{1}/x]\land e_{1}\leq
e_{2}}$, imposes that the constraints collected before executing the
loop are strong enough to imply the invariant right before the start
of the loop.  The configuration to which the for-loop with invariant
steps to has a set of constraint $\cstr_f$ which records the fact that
the for-loop has terminated and so includes the constraint
$\tocstr{\mem_f}{I[e_2+1/x]}$. The final memory $\mem_f$ maps to fresh
symbolic values all the variables, or array names which might have
been updated in the body $c_b$ ($\updateC{\cdot}$ performs a syntactic
check on $c_b$, soundly approximating the set of variables updated by
$c_b$). Notice that we don't update the length of the arrays,
because we consider only arrays of fixed (static or concrete)
length. At the exit of the loop the counter variable has to map to the
value to which the second guard of the for-loop was reduced to.
\subsection{\rslang}
\label{sec:rsfor}
Similarly to what we did in the previous section, we now extend the
language \rlang to \rslang using symbolic values $X$.  The symbolic
extension of the relational language follows the same steps as the
unary with the difference that now symbolic values can also appear in
pairs of expressions $\pairExpr{e_1}{e_2}$ and pairs of commands
$\pairCmd{c_1}{c_2}$ and pairs of values in a memory $(v_1, v_2)$.  As
in the case of the previous languages we give the semantics to
\rslang by means of a big step semantics for symbolic relational
expressions proving judgments of the shape $\seconf{\mem}{e}{\cstr}
\rsestep \send{v}{\cstr'}$, and a small step semantics for
symbolic relational commands proving judgments of the shape
$\sscconf{\mem}{c}{\cstr}\rsscmstep \sscconf{\mem'}{c'}{\cstr'}$.
We provide a selection of the rules to
prove those judgments in Figure \ref{fig:sem-rsfor}.  Projection functions
need now to be smartly extended to relational assertions, this would
be particularly useful for example when a for-loop with invariant $I$
appears in one of the branches of an {\tt if} construct with a guard
which evaluates to a relational value $\pairExpr{v_1}{v_2}$, since both cases
$v>0, v\leq 0$ have to be considered. For this reason we extend
projection functions for basic relational assertions in the following
way (where $\{p,q\}=\{a,b\}$, and where the function $\idx{\cdot}$
returns the set (potentially empty) of indices $i\in\{1,2\}$ appearing
in a relational expression):
\[
  \begin{array}{rcll}  
    \proj{i}{e_a\otimes e_b} &=&\proj{i}{e_a}\otimes\proj{i}{e_b}& \text{if}\  \idx{e_q}\subseteq\idx{e_p}= \{i\}\ \text{or}\\ &&& \idx{e_q}=\idx{e_p}=\emptyset\\
    \proj{i}{e_a\otimes e_b} &=&\true& \text{otherwise}\\
    \proj{i}{x_i}&=&x\\
    \proj{i}{x}&=&x
  \end{array}
\]
For other forms of assertions projection functions behave homomorphically. So
for $i\in\{1,2\}$ we can now define
$\proj{i}{\cforinv{x}{e_1}{e_2}{c_b}{I}}\equiv\cforinv{x}{\proj{i}{e_1}}{\proj{i}{e_2}}{\proj{i}{c_b}}{\proj{i}{I}}$.
Also, $\mergesymmem{\cdot}{\cdot}{\cdot}$ plays a similar role in the
relational symbolic semantics to what $\mergemem{\cdot}{\cdot}{\cdot}$
does in the concrete one.  The rule $\rulestyle{r-s-lift}$ relies on
\slang. It evaluates a relational symbolic expression and returns a single symbolic value
if the two unary symbolic execution reduce to the same integer value, otherwise it splits.
Rule \rulestyle{r-s-arr-ass-split} takes care of an array assignment when the array is
symbolic unary but the right hand side of the assignment is different, and hence the array needs to be split.
Rule \rulestyle{r-s-if-false-true} is similar to the
analogous rule for the concrete semantics we presented in
Figure~\ref{fig:sem-rlang}, the main difference is that now the path
conditions are recorded in the constraint set.
Rule \rulestyle{r-s-if-right} takes care of a pair command with
a branching instruction on the right and a different command on the left.
This rule, and a similar one for the left execution, helps synchronization of the two runs.
In rule \rulestyle{r-s-pair-step} takes care of the general case, where $c_1 \equiv c_2$ means
structural equality, for instance $c_1$ and $c_2$ are both assignments.
Similarly to the analogous concrete rule, one side of the
two is chosen non-nondeterministically, and one step on that side is
performed using the unary symbolic semantics. Finally, the rule
\rulestyle{r-s-for-inv} allows the user to specify a relational
invariant for a for-loop which might diverge because one of the guards
evaluates to a value containing a symbolic value. The rule \rulestyle{r-s-for-inv}
behaves similarly to \rulestyle{s-for-inv} but in a relational setting.
\begin{figure}
  \begin{mathpar}    
      \inferrule[\rulestyle{r-s-lift}]
      {
        \seconf{\proj{1}{\mem}}{\proj{1}{e}}{\cstr}\sestep \send{v_{1}}{\cstr'}
        \and
        \seconf{\proj{2}{\mem}}{\proj{2}{e}}{\cstr'}\sestep \send{v_{2}}{\cstr''}
        \\\\
        \rsend{v}{\cstr'''}=\left \{
          \begin{array}{rcl}
            \rsend{v_{1}}{\cstr''}  &\text{if}& (v_{1},v_{2})\in\forvalset^{2} \land v_{1}=v_{2}\cr
            \rsend{(v_{1}, v_{2})}{\cstr''}  &&\text{otherwise}
          \end{array}
      \right .\
      }
      {
        \rsstart{\mem}{e}{\cstr}\rsestep\rsend{v}{\cstr'''}
      }

  \inferrule[\rulestyle{r-s-arr-ass-split}]
    {
      \mem(a)=(X,l)\in\symvalset\times\sforvalset
      \and
      \fresh{Z}
      \and
      \fresh{W}
      \\\\
      \rsstart{\mem}{e_{i}}{\cstr}\rsestep\rsend{v_{i}}{\cstr'}
      \and
      \rsstart{\mem}{e_{h}}{\cstr'}\rsestep\rsend{v_{h}}{\cstr''}
      \\\\
      \cstr'''=\cstr'\cup\{\proj{1}{v_{h}}\neq \proj{2}{v_{h}}, Z=\store{X}{\proj{1}{v_{i}}}{\proj{1}{v_{h}}}\}\\\\
      \cstr''''=\cstr'''\cup\{W=\store{X}{\proj{2}{v_{i}}}{\proj{2}{v_{h}}},0<\proj{1}{v_{i}}\leq l,0<\proj{2}{v_{i}}\leq l\}
    }
    {
      \rsconf{\mem}{\ass{\arracc{a}{e_{i}}}{e_{h}}}{\cstr}\rsscmstep\rsconf{\mem[a\mapsto ((Z,l),(W,l))]}{\skipp}{\cstr''''}
    }

    \inferrule[\rulestyle{r-s-if-false-true}]
    {\rsstart{\mem}{e}{\cstr}\rsestep\rsend{v}{\cstr'}
      \\\\
      \cstr''=\cstr'\cup\{\proj{1}{v}\leq 0, \proj{2}{v}>0\}
    }
    {\rsconf{\mem}{\ifte{e}{c_{\mathit{tt}}}{c_{\mathit{ff}}}}{\cstr}\rsscmstep
      \rsconf{\mem}{\pairCmd{\proj{1}{c_{\mathit{ff}}}}{\proj{2}{c_{\mathit{tt}}}}}{\cstr''}}
    
 \inferrule[\rulestyle{r-s-pair-step}]
    {
      \sscconf{\proj{i}{\mem}}{c_{i}}{\cstr}\sscmstep\sscconf{\mem'_{i}}{c'_{i}}{\cstr''}
      \\\\
      \big ( \ifte{\cdot}{\cdot}{\cdot}\neq c_{j}=c'_{j}\and \textup{or}\and c_1\equiv c_2 \bigg )
      \\\\
      \{1,2\}=\{i,j\}
      \and
      \mem'_j=\proj{j}{\mem}
      \and
      \mem'=\mergesymmem{i}{\mem'_{1}}{\mem'_2}
    }
    {\rsconf{\mem}{\pairCmd{c_{1}}{c_{2}}}{\cstr}\rsscmstep\rsconf{\mem'}{\pairCmd{c'_{1}}{c'_{2}}}{\cstr''}}

     \inferrule[\rulestyle{r-s-if-right}]
    {
      c_1\equiv \ifte{\cdot}{\cdot}{\cdot} \and c_2\notin\{\ifte{\cdot}{\cdot}{\cdot},\skipp\}
      \\\\
      \sscconf{\proj{2}{\mem}}{c_{2}}{\cstr}\sscmstep\sscconf{\mem'_{2}}{c'_{2}}{\cstr''}\and
      \mem'=\mergesymmem{i}{\proj{1}{\mem}}{\mem'_2}
    }
    {\rsconf{\mem}{\pairCmd{c_{1}}{c_{2}}}{\cstr}\rsscmstep\rsconf{\mem'}{\pairCmd{c_{1}}{c'_{2}}}{\cstr''}}

    \inferrule[\rulestyle{r-s-for-inv}]
    {
      \rsstart{\mem}{e_{a}}{\cstr}\rsestep\rsend{v_{a}}{\cstr'}
      \and
      \rsstart{\mem}{e_{b}}{\cstr'}\rsestep\rsend{v_{b}}{\cstr''}
      \\\\
      \vDash\triple{I\land e_{a}\leq x\land x\leq e_{b}}{c}{I[x_{1}+1/x_{1}][x_{2}+1/x_{2}]}\\\\
      \models \cstr''\Rightarrow\tocstr{\mem}{I[\proj{1}{e_{a}}/x_{1}][\proj{2}{e_{a}}/x_{2}]\land e_a\leq e_b}
      \\\\
      \cstr_{f}=\cstr''\cup\{\tocstr{\mem_{f}}{I[\proj{1}{v_{b}}+1/x_{1}][\proj{2}{v_{b}}+1/x_{2}]}\}
      \\\\
      \hspace{-5pt}\mem_{f}=\lambda n.\left \{
        \begin{array}{rcl}
          v_{b}, && \text{if}\ n=x \cr
                    (X,Y), && \text{if}\ n\in\updaterC{c}, \mem(n)\in\sforvalset\cup\sforvalset^2\cr
                              &&\fresh{X},\fresh{Y}\cr
                                 ((X,l),(Y,l)), && \text{if}\ n\in\updaterC{c}, \mem(n)\in \sarrayset\cr
                              &&\pi_2(\mem(n))=l\cr
                                  ((X,l),(Y,l)), && \text{if}\ n\in\updaterC{c}, \mem(n)\in \sarrayset^2\cr
                              &&\pi_2(\pi_2(\mem(n)))=l\cr
          \mem(n),\ && \text{otherwise}          
        \end{array}
      \right .\
      \\\\
    }
    {\rsconf{\mem}{\cforinv{x}{e_{a}}{e_{b}}{c}{I}}{\cstr}\rsscmstep\rsconf{\mem_{f}}{\skipp}{\cstr_{f}}}
  \end{mathpar}
     \caption{Semantics of \rslang (simplified selected rules).}
    \label{fig:sem-rsfor} 
  \end{figure}
  \subsection{Unary and relational collecting semantics}
  \label{sec:setsems}
  Building on the $\sscmstep$ and $\rscmstep$ semantics, we define now
two collecting semantics which consider only reachable
configurations, namely those whose set of constraints is satisfiable.
Overloading the symbol $\Rightarrow$ we will denote by it both the
unary and relational collecting semantics. Both semantics
are defined through only one rule presented in Figure
\ref{fig:set-sem}.  In rule \rulestyle{set-step} we remove from the
set of configurations taken in consideration the current configuration
and we add to it all the configurations reachable in one step that are
satisfiable.
\begin{figure}[H]
    \begin{mathpar}
       \inferrule[\rulestyle{set-step}]
      {
        \setconf_t=\{\sscconf{\mem'}{c'}{\cstr'} \mid \sscconf{\mem}{c}{\cstr}\xrightarrow{\dagger}\sscconf{\mem'}{c'}{\cstr'} \land \sat{\cstr'}\}\\\\
         \sscconf{\mem}{c}{\cstr} \in \setconf\and \setconf'=\bigg(\setconf\setminus\{\sscconf{\mem}{c}{\cstr}\}\bigg ) \cup \setconf_t 
      }
       {
        \setconf \Rightarrow \setconf'
      }
    \end{mathpar}
    \caption{Unary and relational collecting semantics rule schema. \\
      $\dagger\in\{\vbox{\hbox{\tiny{\textsf{\textup{SF}}}}}, \vbox{\hbox{\tiny{\textsf{\textup{RSF}}}}}\}$}
    \label{fig:set-sem}
  \end{figure}
  
  \section{Meta theory}
  \label{sec:metatheory}
  In this section we will make more precise the connection between  concrete and symbolic languages.
  In order to do this, we need to reason about \emph{ground
    substitutions} turning object containing symbolic values into
  concrete objects.   Given a command $c$ or an expression $e$ in \slang (or in \rslang) and a ground substitution $\sigma \in\Sigma$
  we write $\sub{c}{\sigma}$ (and $\sub{e}{\sigma}$)
  for the application of $\sigma$ to $c$ (and $e$). 
  We can also apply a substitution to a unary symbolic memory:
  \begin{mydef}
  Given a ground substitution $\sigma\in\subs$ we define its application to a unary symbolic  memory  as 
  \begin{center}
    $    \sub{\mem}{\sigma}=\lambda m.
    \left \{
      \begin{array}{rcl}
        \sigma(\mem(m)) && \text{if}\ \mem(m)\in\symvalset \cr
        \sigma(\pi_{1}(\mem(m)))  && \text{if}\ \mem(m)\in \symvalset\times \textup{\sforvalset} \cr
        \mem(m) && \text{otherwise}
      \end{array}
    \right .\
    $
  \end{center}
  where $m$ ranges over $\varset\cup\vararrset$.
\end{mydef}
We have a similar definition for relational symbolic memories which we
omit here.
From now on, we consider only substitutions
$\sigma$ which respect the type of the program variables and array
names appearing in a symbolic expression or command. That is, given an expressing $e$ (or command $c$) 
we consider substitutions $\sigma$ for which $\sub{e}{\sigma}$ ($\sub{c}{\sigma}$) is
an expression (command) in \lang (\rlang) whenever $e$ ($c$) is an
expression (command) in \slang (\rslang).  
We also want to consider only substitutions mapping 
symbolic values to objects of their type. This is characterized by the
following definition. 
\begin{mydef}{\label{def:validation}} We say that a ground
substitution $\sigma\in\subs$ validates a configuration $
\sscconf{\mem}{c}{\cstr}$ and we write $\validatec{\sigma}{\sscconf{\mem}{c}{\cstr}}$ iff $\sigma \models\cstr$, $\forall
a\in\vararrset. \mem(a)=(X,v)\Rightarrow \sub{X}{\sigma}\in \{1,\dots,{\sub{v}{\sigma}}\}\rightarrow \forvalset$, $\forall x\in\varset. 
\mem(x)=X \Rightarrow \sub{X}{\sigma}\in\forvalset$, and
$\sigma$ respects the type of array names and program variables in
$c$.
\end{mydef}
We also consider the natural partial order, $\preceq$ over
$\subs$ given by the relation $\{(\sigma_1,\sigma_2)\in\subs^2\mid \forall X
\in\dom{\sigma_1}. \sub{X}{\sigma_1}=\sub{X}{\sigma_2}\}$ .

\subsection{Coverage }
We now want to formalize the idea that a run of the set semantics can
capture (\emph{cover}) many concrete runs. To do this we formalize
what a \emph{final} configuration ($\setconf$) of the $\Rightarrow$
semantics (Figure \ref{fig:set-sem}) is.
  \begin{mydef} A unary (or relational) configuration $s$ is final, 
 and we write  $\final{s}$, when $s=\sscconf{\mem}{\skipp}{\cstr}$.
A set of configurations $\setconf$ is final, denoted
$\final{\setconf}$, if and only if forall $s \in \setconf. \final{s}$.
  \end{mydef} 
The following lemma states that any concrete execution
  can be covered by a symbolic path. This symbolic path will have
  a satisfiable set of constraints which will make it  possible to map
  back symbolic final configurations to
  the concrete final configuration of the concrete path.
\begin{lemma}\label{lem:coverage}
    If $\setconf\Rightarrow^{*}\setconf'$,
     $\sscconf{\mem_1}{c_1}{\cstr_1}\in\setconf$, and 
     $\validatec{\sigma_1}{ \sscconf{\mem_1}{c_1}{\cstr_1}}$
     then 
     $\exists k_{c_{1}},\exists\sscconf{\mem_2}{c_2}{\cstr_2}\in\setconf',\exists \sigma_2\in\subs$ such that
     $\scconf{\sub{\mem_1}{\sigma_1}}{c_1}\scmstep^{k_{c_{1}}}\scconf{\sub{\mem_2}{\sigma_2}}{c_2}$
      (or $~\scconf{\sub{\mem_1}{\sigma_1}}{c_1}\rscmstep^{k_{c_{1}}}\scconf{\sub{\mem_2}{\sigma_2}}{c_2}$), 
     $\sigma_2\vDash\sscconf{\mem_2}{c_2}{\cstr_2}$, and 
     $\sigma_1 \preceq \sigma_2$.
  \end{lemma}
\subsection{Proving and soundness}
  In symbolic execution we want to execute symbolically a program in
  order to reason about multiple concrete executions. In order to do
  this we need to specify an initial memory from which the symbolic
  execution can start. Without loss of generality we choose as initial
  memory the most abstract. This leads to the following definition:
\begin{mydef}
  \label{def:abstract-mem}
  Let $\Phi$ be a unary assertion, and $c$ a command in
\lang.  Define the following symbolic memory:
$$\absmem{\Phi,c}\equiv\lambda n\in\varof{\Phi}\cup\varof{c}.\left \{
        \begin{array}{rcl} X, && \text{if}\ n\in\varset \cr (X, L), &&
\text{if}\ n\in\vararrset \cr
        \end{array} \right .\ $$
 where all the variables $X,L$ are meant to be
distinct and fresh, and the function $\varof{\cdot}$ returns the set
of program variables and array names appearing in the argument.
\end{mydef}
The previous definition can be easily extended to relational memories,
assertions, and commands.  
As we already discussed, we are interested in using \rse
for proving valid specifications of programs. If
we want to prove that a triple $\triple{\Phi}{c}{\Psi}$ is valid, we
can execute symbolically $c$ starting from an initial symbolic
configuration which satisfies the precondition $\Phi$. If we 
reach only final configurations whose set of constraints
imply the postcondition $\Psi$,
then the triple is valid.  Formally:
\begin{mydef}
  Let $c$ be a command in \lang (or \rlang) and $\Phi$ and $\Psi$ unary (or
relational) assertions. We say that $c$ symbolically proves $\Psi$ from
$\Phi$, and we write $\provable{\Phi}{c}{\Psi}$ iff there exists $\setconf$ such that
  \begin{itemize}
  \item
$\{\sscconf{\absmem{\Phi,c}}{c}{\{\tocstr{\absmem{\Phi,c}}{\Phi}\}}\}\setsemantics^{*}\setconf$
  \item $\final{\setconf}$
  \item $\forall
\sscconf{\mem}{\skipp}{\cstr}\in\setconf. \models \cstr\implies \tocstr{\mem}{\Psi}$
  \end{itemize}
\end{mydef}
It now makes sense to formulate the following soundness theorem:
\begin{theorem}[Soundness of verification]
  \label{thm:soundness-verification}
   Let $\Phi$ and $\Psi$ be unary (or relational) assertions and  let $c$ be
   a command in \lang (or \rlang). Then, if $\provable{\Phi}{c}{\Psi}$ then $\vDash\triple{\Phi}{c}{\Psi}$.
   \begin{proof} By structural induction on $c$, using Lemma \ref{lem:coverage}.\end{proof}
  \end{theorem} 
  \subsection{Finding counterexamples: strength of invariants and soundness}
  \label{subsec:strength}
 We now want to formalize the fact that we can use \rse for finding
 counterexamples.
Let us consider a program $c$, a precondition $\Phi$ and a postcondition $\Psi$.  If
starting to evaluate $c$ from an initial symbolic configuration and a set of
constraints that satisfy the precondition $\Phi$, we arrive in a final configuration
whose set of constraint is consistent with the negation of the
postcondition $\Psi$ (interpreted in the memory of the final configuration),
then we know that the post-condition does not hold. This argument motivates the following definition.

  \begin{mydef}
    \label{def:disproves}
    Let $c$ be a command in \lang (or \rlang) and $\Phi,\Psi$ unary (or relational) assertions.
    We say that, $c$ symbolically disproves $\Psi$ from $\Phi$
    and we write $\disproves{\Phi}{c}{\Psi}$ if and only if   exists $\setconf$ such that
  \begin{itemize}
  \item $\{\sscconf{\absmem{\Phi,c}}{c}{\{\tocstr{\absmem{\Phi,c}}{\Phi}\}}\}\setsemantics^{*}\setconf$
  \item $\exists \sscconf{\mem}{\skipp}{\cstr}\in\setconf. \sat{ \cstr\cup\{\tocstr{\mem}{\neg \Psi}\}}$
  \end{itemize}
\end{mydef}

A counterexample to the validity of a unary triple $\triple{\Phi}{c}{\Psi}$ consists of a pair of
concrete memories $\mem_1,\mem_2$ and $\intlog\in\intlogs$ such that $\mem_1\vDash_{\intlog} \Phi$ and
$\scconf{\mem_1}{c}\scmsteptr\scconf{\mem_2}{\skipp}$ but
$\mem_2\nvDash_{\intlog} \Psi$.

We would like to be able to extract, from an execution showing $\disproves{\Phi}{c}{\Psi}$, a
counterexample for $\triple{\Phi}{c}{\Psi}$. Unfortunately, this cannot 
always be done. 

Indeed because in presence of loops, invariants might just approximate the
state after the loop has terminated.
That is the invariants might not specify precisely enough
the state after the loop body has been executed $n$ times for arbitrary $n$.
For instance: $\triple{z=0 \land x>0}{\cforinv{i}{1}{x}{~~\ass{z}{z+1}}{\true}}{\false}$ is obviously an
invalid triple but the invariant does not say
much about the value of $z$ after the loop has been executed $x$ times.
The invariant $I_s\equiv z=i$ would instead do the job, specifying exactly the final state.
When invariants have this property we say they are \emph{strong}.
With Definition
\ref{def:strength-inv} we capture the notion of \emph{strength} of an
invariant. 
  \begin{mydef}
    \label{def:strength-inv}
  Given a command $c\equiv\cforinv{x}{e_1}{e_2}{c_b}{I}$ in \lang (or in \rlang), we say that the invariant $I$ is strong iff $~\forall \sigma_{1},\sigma_{2}\in\subs$,
  if $\validatec{\sigma_{1}}{\sscconf{\mem_f}{\skipp}{\cstr_{f}}}$,
  $\validatec{\sigma_{2}}{\sscconf{\mem_f}{\skipp}{\cstr_{f}}}$, and
  $\sub{\mem}{\sigma_{1}}=_{R}\sub{\mem}{\sigma_{2}}$
  then $\sub{\mem_{f}}{\sigma_{1}}=_{U}\sub{\mem_{f}}{\sigma_{2}}$.

  Where $R=\bigg((\varset\cup\vararrset)\setminus U\bigg)\cup\{x\}$,
  $U=\updateC{c_{b}}$ (or $U=\updaterC{c_b}$), and $\mem,\mem_f,\cstr_f$ are
  respectively the memory right before the execution of the for-loop, and 
  the memory and the set of constraints after the application of the rule \rulestyle{s-for-inv} (or \rulestyle{r-s-for-inv}).
\end{mydef}

The following theorem allows to avoid false positives in interactive refutation.
\begin{theorem}[Soundness of counterexample finding]
  \label{thm:sound-bug-finding}
  Let $\Phi,\Psi$ be unary (or relational) assertions and $c$ a command in
  \lang (or \rlang). Then,  if $\disproves{\Phi}{c}{\Psi}$ and all the invariants in $c$ (if any) are strong then 
  $\not\vDash\triple{\Phi}{c}{\Psi}$.
\end{theorem}
Theorem \ref{thm:sound-bug-finding} is a soundness result for
counterexample finding which implies (relative) completeness of the
proving system w.r.t to the semantics of $\lang$ (and $\rlang$). Indeed,
provided the program $c$ is annotated with strong enough invariants, if
\rse cannot derive $\disproves{\Phi}{c}{\Psi}$ then it has to be the case
that $\vDash\triple{\Phi}{c}{\Psi}$. The completeness just mentioned
concerns the proving system and has nothing to do
with the semantic completeness of $\rlang$ w.r.t to $\lang$ which has been
already ruled out in Section \ref{sec:relation-assert}.
\section{Implementation}
\label{sec:implementation}
\rse has been implemented in OCaml 4.06 in about 4k LOC.  The queries
on satisfiability of set of constraints are discharged using the SMT
solver Z3~\cite{DeMoura}. The implementation is not fully optimized.
\subsection{Checking the semantic judgment}
Rules \rulestyle{s-for-inv} and \rulestyle{r-s-for-inv}, include among the
premises a Hoare triple validity judgment, which ensures that
the assertion provided is an inductive invariant of the loop. By using
semantic validity we allow other potential implementations to
use different analysis techniques for the verification  of that triple, e.g., a sound
Hoare logic for \lang (or \rlang).  Since we want \rse to be  a
self-contained tool, in the implementation we prove this judgment by
recursively calling \rse.  In particular, while
executing the rule \rulestyle{s-for-inv} (or \rulestyle{r-s-for-inv}) on the command
$\cforinv{x}{e_1}{e_2}{c}{I}$ we use recursively \rse to prove
$\vDash\triple{I\wedge e_1\leq x \wedge x\leq e_2}{c}{I[x+1/x]}$, by checking that
indeed $\provable{I\wedge e_1\leq x \wedge x\leq e_2}{c}{I[x+1/x]}$.
This can
also help in practice in finding the \emph{right} invariant by giving
the user prompt feedback on why the assertion used at the moment
is not an inductive invariant.
\subsection{Checking the strength of the invariant}
If we want to use \rse for finding counterexamples to specifications, we might need to check that the invariant is
strong as in Definition (\ref{def:strength-inv}), so that by Theorem
\ref{thm:sound-bug-finding} we can be sure that the ground
substitution provided (if any) by the SMT is indeed a counterexample. 
In particular,  this ensures that, if the SMT returns a
$\sigma$ such that $\sigma\models \cstr_f\cup\{\neg\tocstr{\mem_f}{\Psi}\}$, then indeed:
\begin{itemize}
\item $\sub{\absmem{\Phi,c}}{\sigma}\vDash  \Phi$
\item  $\scconf{\sub{\absmem{\Phi,c}}{\sigma}}{\sub{c}{\sigma}} \scmsteptr\scconf{\sub{\mem_f}{\sigma}}{\skipp}$
      \\(or $\scconf{\sub{\mem_1}{\sigma_1}}{c_1}\rscmstep^{*}\scconf{\sub{\mem_2}{\sigma_2}}{c_2}$), 
\item $\sub{\mem_f}{\sigma}\vDash \neg \Psi$
\end{itemize}
A way to check this property is to check for unsatisfiability the
following  set of constraints:
\[
  \cstr_f\cup\{ \tocstr{\mem_{f}}{I[v_2+1/x]}^{\mathbf{F}} \}  \cup\{\bigvee_{\{X\in \mathbf{F}\}}X\neq X'\}
\]
where: $\mathbf{F}$ is the set of fresh
symbols generated during the execution of the rule \rulestyle{s-for-inv} (or \rulestyle{r-s-for-inv}),
and $ \tocstr{\mem_f}{I[v_2+1/x]}^{\mathbf{F}}$ is the result of taking the invariant 
where $x$ has been substituted with $v_2+1$, interpreted as a constraint through $\mem_f$,
with all the symbols in $\mathbf{F}$ substituted with their primed
versions. If it is not the case that $\sat{ \cstr_f\cup\{ \tocstr{\mem_{f}}{I[v_2+1/x]}^{\mathbf{F}} \}  \cup\{\bigvee_{\{X\in \mathbf{F}\}}X\neq X'\}}$ then
there is only a possible way to satisfy $\cstr_f$ once the symbols generated before the loop have been fixed,
that is given a ground substitution
$\sigma$ for which $\sigma\vDash \cstr''$  then there is only one possible $\sigma'$ such that
$\sigma\preceq\sigma'$ and $\sigma'\vDash\cstr_f$. This implies the strength of the invariant $I$.
\section{Experimental results}
\label{sec:experiments}
We compared our relational symbolic semantics with other techniques
used to prove or finding counterexamples to relational properties. In
particular with naive self-composition, simple product
programs and the product programs construction of~\cite{Eilers18}.
Since our implementation does not use
any heuristics to try to improve efficiency it makes sense to compare it
with vanilla versions of these techniques. Also, notice that
product programs and self-composition can
be easily embedded in our framework by just executing self-composed
programs and product programs in \slang, that is by just using unary
symbolic semantics. In this section we can
see some experimental results that show that relational symbolic
execution is comparable in terms of execution time, calls to the
solver, and number of steps with respect to self-composition and
product programs.
The results in Table (\ref{tab:exp-results-proving}) are
about proving relational properties, while in Table (\ref{tab:exp-results-disproving}) the results
are about finding counterexamples to relational properties.
Some of the examples are taken from standard literature (sometimes
adapting them to our language).
In the table an $R$ (Relational) means that relational
symbolic execution was used, while $U$ denotes that the self composed program
was analyzed with unary symbolic 
semantics, a P denotes a product program symbolically executed with unary semantics.
Because of space reasons we only show information which showed
discernible differences in resource usage. 
An $\uparrow$ denotes that
the symbolic execution had to be terminated because it was running for
too long, while an \xmark{} means that the SMT solver was not able to
discharge a query and so the result is unknown. Finally, a $?$ denotes
absence of information, necessary when \rse{} ran out of time limits.
The results regarding execution time are
an average over 50 runs executed on an Intel CPU, 2.80GHz with 16 GB
of RAM memory. 

The examples concern properties such as
non-interference, e.g.,~\emph{n-inter. example} and \emph{inter. example} series, \emph{ni-array} example,
or execution time independence e.g.,~\cite{Antonopoulos}, or continuity e.g.,~\emph{sum-k-lip-cont}, or ~\emph{sort-k-lip-cont}.
\begin{table}
  \centering
\begin{tabular}{lllllll}
  \hline
  Example&R/U/P&\#BS&\#SS& \#SMT &\#S & tm (s)\\
  \hline
  \hline
  Darvas   &\vline R&\vline     8&\vline                 5&\vline                       3&\vline             2&\vline                 0.39\\
  Darvas   &\vline U&\vline    15&\vline                18&\vline                     5&\vline             3&\vline                 0.04\\
  Costanzo &\vline     R&\vline     30032&\vline            42315&\vline                   10921&\vline         4096&\vline              139\\
  Costanzo &\vline     U&\vline     ?&\vline            ?&\vline                   ?&\vline         ?&\vline              $\uparrow$\\
  Antonopoulos&\vline        R&\vline     68&\vline                101&\vline                     20&\vline            10&\vline                0.15\\
  Antonopoulos&\vline        U&\vline    70&\vline                 94&\vline                      22&\vline            10&\vline                0.16\\
  Terauchi[1]&\vline     R&\vline     34&\vline                63&\vline                      24&\vline            4&\vline                 0.22\\
  Terauchi[1]&\vline     P&\vline     309&\vline                2472&\vline                      179&\vline            9&\vline                 1.49\\
  Terauchi[1]&\vline       U&\vline    46&\vline                 141&\vline                     22&\vline            4&\vline                 0.22\\
  Terauchi[2]&\vline      R&\vline     55&\vline                91&\vline                      34&\vline            9&\vline                 0.36\\
  Terauchi[2]&\vline      U&\vline     31&\vline                155&\vline                      10&\vline            3&\vline                 \xmark\\
  n-inter. example 1&\vline            R&\vline     5&\vline                 4&\vline                       1&\vline             1&\vline                 0.01\\
  n-inter. example 1&\vline            P&\vline     33&\vline                 56&\vline                       26&\vline             4&\vline                 0.01\\
  n-inter. example 1&\vline            U&\vline    9&\vline                  16&\vline                      1&\vline             1&\vline                 0.01\\
  n-inter. example 2&\vline            R&\vline     5&\vline                 4&\vline                       1&\vline             1&\vline                 0.01\\
  n-inter. example 2&\vline            U&\vline    9&\vline                  16&\vline                      1&\vline             1&\vline                 0.01\\
  n-inter. example 3&\vline            R&\vline     7&\vline                 9&\vline                       1&\vline             1&\vline                 0.01\\
  n-inter. example 3&\vline            P&\vline     30&\vline                 87&\vline                       24&\vline             2&\vline                 0.16\\
  n-inter. example 3&\vline            U&\vline    13&\vline                 36&\vline                      1&\vline             1&\vline                 0.02\\
  n-inter. example 4&\vline            R&\vline     14&\vline                13&\vline                      8&\vline             4&\vline                 0.08\\
  n-inter. example 4&\vline            P&\vline     51&\vline                 109&\vline                    35&\vline          9&\vline                  0.2\\
  n-inter. example 4&\vline            U&\vline    16&\vline                 14&\vline                      10&\vline            4&\vline                 0.1\\
  sum-k-lip-cont&\vline            R&\vline     8&\vline                 5&\vline                       3&\vline             2&\vline                 0.04\\
  sum-k-lip-cont&\vline            U&\vline     8&\vline                 11&\vline                       3&\vline             2&\vline                 \xmark\\
  sort-k-lip-cont&\vline             U&\vline    55&\vline                72&\vline                       23&\vline             12&\vline                0.12\\
  sort-k-lip-cont&\vline             R&\vline    31&\vline              45&\vline                      12&\vline            6&\vline                0.11\\
  sort-k-lip-cont&\vline             P&\vline    50&\vline               66&\vline                      15&\vline           12&\vline              0.12\\
  \hline
\end{tabular}
\caption{Experimental results of proving relational properties. Where Darvas stands for \cite{Darvas}, Costanzo stands for \cite{Costanzo},
Antonopoulos stands for \cite{Antonopoulos}, Terauchi stands for \cite{Terauchi}}
\label{tab:exp-results-proving}
\end{table}
\begin{table}
\centering
  \begin{tabular}{lllllll}
  \hline
  Example&R/P/U&\#BS&\#SS& \#SMT &\#S & tm (s)\\
  \hline
  \hline
  ni-array&\vline        R&\vline       291&\vline             380&\vline               132&\vline                   37&\vline        1.08\\ 
  ni-array&\vline        U&\vline       342&\vline             674&\vline               90&\vline                    16&\vline        0.7\\
  \cite{Eilers18}&\vline        R&\vline       9&\vline               7&\vline                 3&\vline                     2&\vline         0.03\\
  \cite{Eilers18}&\vline        U&\vline       13&\vline              25&\vline                1&\vline                     1&\vline         0.01\\
  inter. example1&\vline     R&\vline       3&\vline               1&\vline               1&\vline                       1&\vline         0.01\\
  inter. example1&\vline     P&\vline       13&\vline               10&\vline               10&\vline                       4&\vline         0.08\\
  inter. example1&\vline     U&\vline       21&\vline              33&\vline              5&\vline                       3&\vline         0.04\\
  inter. example2&\vline     R&\vline       3&\vline               1&\vline               1&\vline                       1&\vline         0.01\\
  inter. example2&\vline     P&\vline       13&\vline               10&\vline               10&\vline                       4&\vline         0.07\\
  inter. example2&\vline     U&\vline       5&\vline               4&\vline               1&\vline                       1&\vline         0.02\\
  inter-password&\vline     R&\vline       485&\vline             714&\vline             169&\vline                     64&\vline        1.40\\
  inter-password&\vline     U&\vline       703&\vline             960&\vline               190&\vline            64&\vline                 1.78\\
  \hline
\end{tabular}
\caption{Experimental results for finding counterexamples relational properties.}
\label{tab:exp-results-disproving}
\end{table}
On this benchmark overall relational symbolic execution performs
better with respect to standard unary self composition and comparably to product
programs, in terms of execution time.  Besides execution
times (unary and relational semantics) we can consider as measures also
other information such as the number of steps of the semantics (small-step \#SS, big-steps \#BS)
performed, calls to the solver (\#SMT) and number of final states 
reached (\#S). Using these metrics shows more clearly how a relational
approach can, at times, outperform other approaches for the
verification or interactive refutation of relational properties.
\cite{Eilers18} construction for product programs introduces
new variables
and new branching instructions. This is
the main reasons why the number of SMT calls increases.  More
generally: consider the base product program construction in
\cite{BartheCrespo} and the number of basic instructions performed
(e.g. assignments) as a measure: commands are duplicated even
when it doesn't help. Product self-composition is a generic
syntactic technique. E.g.: take $c\equiv \ass{p}{p+1}$, and suppose we want to
show that: $\vDash\triple{p1=p2}c{p1=p2}$. Under product programs we could
reduce the problem to verifying: $\vDash \triple{p1=p2}{c1 \times c2}{p1=p2}$ that is
$\vDash \triple{p1=p2}{\seq{\ass{p1}{p1+1}}{\ass{p2}{p2+1}}}{p1=p2}$.
In the unary symbolic execution of the product program \emph{necessarily} two assignments will
be performed. While executing relationally $\ass{p}{p+1}$ \emph{might} only
execute one assignment.

This evaluation shows that although we have trade-offs between the
different techniques and none of them is always better, in several
situations relational symbolic execution brings clear improvements.

\section{Related Works}
  \label{sec:related-works}
  The works most closely related to ours are the ones that have used
  symbolic execution for relational properties. ~\cite{Milushev} use symbolic execution to check non-interference by means
  of an analysis based on a type directed transformation of the
  program first presented in ~\cite{Terauchi}. The analysis targets programs written in a subset
  of C which includes procedures calls, and dynamically allocated data
  structures modeled through a heap. A main difference with our work
  is that they focus only on non-interference while we focus on arbitrary
  relational properties. Additionally, they use self-composition while
  we focus on the design of a formal relational semantics. Finally,
  they use a generic approach based on heaps, instead, we focus on
  arrays as concrete data structures and we leverage their properties
  in the design of our semantics.

  In \cite{Person} symbolic execution is used to check differences
  between program versions. The property they analyze,  although relational can be
  easily described with two separate execution of the two programs.
  Indeed, in their work symbolic execution is used separately for the two programs.

  Relational properties have also been studied through many other
  techniques. 
  We already mentioned different works that reduce the verification of
relational properties to the one of properties through
self-composition~\cite{Barthe,Terauchi} and product
programs~\cite{BartheCrespo,Asada, Eilers18}.  Several works have
studied relational versions of Hoare logics. For example,
Benton~\cite{Benton} studies relational Hoare logics for
noninterference and program equivalence, and Barthe et
al.~\cite{Barthe2012,Barthe2014} study relational Hoare logics for
relational probabilistic properties, such as differential privacy. Their work is based
on a denotational semantics based on couplings and probabilistic liftings, while ours
is operational in nature. Other works such
as~\cite{BanerjeeNN16} have focused on a relational Hoare logics with
frame rules to deal with heap based semantics, and on situations where
keeping the traces not aligned might be beneficial in the same spirit
of dissonant loop rules introduced in \cite{Beringer2011}. Other works
instead tried to maximize the amount of synchronicity between the two
runs~\cite{Pick2018}.
  Several works have studied type systems for the verification of
  different relational properties, some examples
  are noninterference~\cite{VolpanoIS96,Pottier,Nanevski}, security of
  cryptographic implementations~\cite{BartheFGSSB14}, differential
  privacy and mechanism design~\cite{Barthe2015}, and relational cost
  \cite{Relcost}
  These approaches are quite different from ours. For instance \cite{Relcost}
  focuses on functional programs, and uses a type discipline
  which requires a lot of domain expertise.
Other works have applied abstract interpretation techniques to 
noninterference~\cite{GiacobazziM04,feret:sas01,AssafNSTT17}. While
symbolic execution and abstract interpretations share several
similarities, the techniques that the approaches rely on are quite
different. 
In ~\cite{AustinFlanagan} authors introduce faceted values,
that resemble our paired values. They do this to simulate simultaneous
runs of the same program on different security levels,
in order to provide information flow security with
a dynamic approach as opposed to a static one as we do in this work.
Cartesian Hoare Logic~\cite{Sousa2016} and its quantitative
extension~\cite{ChenFD17} can be used for reasoning about
generic k-safety properties, and their quantitative analogous. 
The language that Cartesian Hoare Logic considers includes arrays and
while loops with breaks. The class of properties they consider goes
beyond relational properties and their analysis is automated. The main
difference between their approach and ours is that we perform symbolic
execution which can also be used to finding bugs while they only focus,
at least on the theoretical part, on proving correctness via Hoare Logic.
Kwon et al.~\cite{KwonHE17}  recently proposed a program analysis for
checking information flow policies over streams based on a technique 
for synthesizing relational invariants. This analysis is not
based on symbolic execution, but we plan to explore if their algorithm 
for synthesizing relational invariants can be used in our setting.

Similar to our work
their semantics is based on couplings and the probabilistic lifting of
relations.  Close to our work is also \cite{albarghouthiH18} where a
proof technique, casting differential privacy proofs as a strategy in
a game encoded as a set of constraints, is presented.  In that work
authors focus again in finding proof and not in finding counter
examples to differential privacy.

\section{Conclusions}
  \label{sec:conclusion}
  In this work we presented \rse, a foundational framework for
  relational symbolic execution. The framework
  supports interactive refutation as well as proving of relational properties for
  a language with arrays and loops.
  We provided some meta theoretical results about symbolic
  execution for its use with respect to proving validity of triples
  and disproving them and we provided necessary conditions
  for which disproving is actually sound. 
  We have shown the flexibility of
  this approach by analyzing examples for a range of different
  relational properties. We compared the analysis
  of this properties using different approaches, i.e.,
  self-composition, product programs and
  relational approach. We have implemented the tool and 
  in the future we plan to address more complex features like functions,
  promises and closures, as well as exploring the generation
  of relational loop invariants~\cite{Qin,Chen,Hoder,Khurshid,KwonHE17}, limiting
  in this way the need for annotations provided by the user.

\bibliography{biblio}

\end{document}

%% file: macros.tex
\newcommand{\lang}{\textsf{\textup{FOR}}\xspace}
\newcommand{\rlang}{{\textsf{\textup{RFOR}}}\xspace}
\newcommand{\slang}{\textsf{\textup{SFOR}}\xspace}
\newcommand{\rslang}{\textsf{\textup{RSFOR}}\xspace}

\newtheorem{mydef}{Definition}

\newcommand{\len}[1]{{\tt len}(#1)}
\newcommand{\skipp}{{\tt skip}}
\newcommand{\seq}[2]{#1{\tt ;}\,#2}
\newcommand{\ifte}[3]{{\tt if}~#1~{\tt then}~#2~{\tt else}~#3}
\newcommand{\cfor}[4]{{\tt for}~(#1~{\tt in~}#2{:}#3)~{\tt do}~{#4}}
\newcommand{\ass}[2]{#1{\tt \leftarrow}#2}

\newcommand{\arracc}[2]{#1[#2]}

\newcommand{\ints}{\mathbb{Z}}

\newcommand{\varset}{{\textup{\textbf{Var}}}}

\newcommand{\carrayset}{{\ensuremath{\textbf{Array}}}}
\newcommand{\sarrayset}{{\ensuremath{\textbf{Array}_{\textbf{s}}}}}     
\newcommand{\vararrset}{\textup{\textbf{Arrvar}}}
\newcommand{\symvalset}{\textup{\textbf{Symval}}}

\newcommand{\forvalset}{\mathbb{Z}}

\newcommand{\sforvalset}{{\ensuremath{\textbf{V}_{\textbf{s}}}}}

\newcommand{\rforvalset}{\ints\cup\ints^2}

\newcommand{\rcarrayset}{\carrayset\cup\carrayset^2}

\newcommand{\sub}[2]{#2 (#1)}

\newcommand{\validatec}[2]{#1\models #2}

\newtheorem{observation}{Observation}

\newcommand{\rulestyle}[1]{{\textbf{\textsc{#1}}}}

\newcommand{\mem}{\mathcal{M}}
\newcommand{\memset}{\textup{\textbf{Mem}}}

\newcommand{\sat}[1]{\textup{\textbf{SAT}}(#1)}

\newcommand{\fresh}[1]{#1~\textbf{\textup{fresh}}}

\newcommand{\econf}[2]{\langle #1,#2\rangle}

\newcommand{\scconf}[2]{(#1,#2)}
\newcommand{\estep}{\Downarrow_{\vbox{\hbox{\tiny{\textsf{\textup{F}}}}}}}
\newcommand{\scmstep}{\xrightarrow{\vbox{\hbox{\tiny{\textsf{\textup{F}}}}\vskip-2pt}}}
\newcommand{\scmsteptr}{\scmstep\!\!^{*}}

\newcommand{\reconf}[2]{\langle #1,#2\rangle}
\newcommand{\rscconf}[2]{(#1,#2)}
\newcommand{\proj}[2]{\lfloor#2\rfloor_{#1}}

\newcommand{\mergemem}[3]{\textup{\textbf{merge}}(#2,#3)} 
\newcommand{\mergesymmem}[3]{\textup{\textbf{merge-s}}(#2,#3)}
\newcommand{\ssebstep}{\Downarrow_{\textup{se}}}
\newcommand{\seconf}[3]{\langle #1,#2,#3\rangle}

\newcommand{\rsstart}[3]{\langle #1,#2,#3\rangle}
\newcommand{\rsend}[2]{\langle #1,#2\rangle}
\newcommand{\sscconf}[3]{(#1,#2,#3)}

\newcommand{\cstr}{\mathcal{S}}
\newcommand{\setconf}{\mathscr{F}}

\newcommand{\rsconf}[3]{(#1,#2,#3)}

\newcommand{\final}[1]{\textbf{\textup{Final}}(#1)}
\newcommand{\dom}[1]{\textbf{\textup{dom}}(#1)}

\newcommand{\select}[2]{\textup{select}(#1,#2)}
\newcommand{\store}[3]{\textup{store}(#1,#2,#3)}

\newcommand{\alen}[1]{\textup{\textbf{len}}(#1)}

\newcommand{\true}{\textup{\textbf{true}}}
\newcommand{\false}{\textup{\textbf{false}}}

\newcommand{\lvar}{\textup{\textbf{Lvar}}}
\newcommand{\intlogs}{\textup{\textbf{Intlog}}}
\newcommand{\intlog}{\mathcal{I}}

\newcommand{\triple}[3]{\{#1\}#2\{#3\}}
\newcommand{\tocstr}[2]{\llbracket #2\rrbracket_{#1}}
\newcommand{\absmem}[1]{\mathcal{M}em_{#1}}

\newcommand{\updateC}[1]{\textup{\textbf{Upd}}(#1)}
\newcommand{\updaterC}[1]{\textup{\textbf{Updr}}(#1)}

\newcommand{\varof}[1]{\textup{\textbf{VarOf}}(#1)}

\newcommand{\subs}{\Sigma}

\newcommand{\pairExpr}[2]{\langle #1\,|\,#2 \rangle}
\newcommand{\pairCmd}[2]{\langle #1\,|\,#2 \rangle}

\newcommand{\idx}[1]{\ensuremath{\emph{Idx}(#1)}}
\newcommand{\provable}[3]{#2: #1\Longrightarrow#3}
\newcommand{\disproves}[3]{#2: #1\centernot\Longrightarrow#3}

\newcommand{\send}[2]{\langle #1,#2\rangle}

\newcommand{\rsscmstep}{\xrightarrow{\vbox{\hbox{\tiny{\textsf{\textup{RSF}}}}\vskip-2pt}}}

\newcommand{\rsestep}{\Downarrow_{\vbox{\hbox{\tiny{\textsf{\textup{RSF}}}}}}}
\newcommand{\sestep}{\Downarrow_{\vbox{\hbox{\tiny{\textsf{\textup{SF}}}}}}}

\newcommand{\sscmstep}{\xrightarrow{\vbox{\hbox{\tiny{\textsf{\textup{SF}}}}\vskip-2pt}}}
\newcommand{\restep}{\Downarrow_{\vbox{\hbox{\tiny{\textsf{\textup{RF}}}}}}}
\newcommand{\rscmstep}{\xrightarrow{\vbox{\hbox{\tiny{\textsf{\textup{RF}}}}\vskip-2pt}}}

\newcommand{\cforinv}[5]{{\tt for}~(#1~{\tt in~}#2{:}#3)~{\tt do}_{#5}~{#4}}
\newcommand{\setsemantics}{\Rightarrow_{\textup{set}}}

\newcommand{\rse}{\textsf{\textup{RelSym}}\xspace}